\documentclass[AMA,STIX1COL]{WileyNJD-v2}
\usepackage{dsfont}
\usepackage[utf8]{inputenc}
\usepackage{amsthm,amsmath}
\usepackage{bm}
\usepackage{bbm}
\usepackage{cancel}
\usepackage{hyperref}
\DeclareMathOperator{\logit}{logit}
\DeclareMathOperator{\expit}{expit}

\articletype{}%

\received{26 April 2016}
\revised{6 June 2016}
\accepted{6 June 2016}

\begin{document}

\title{Transportability of model-based estimands in evidence synthesis}

\author[1]{Antonio Remiro-Az\'ocar}

\authormark{REMIRO-AZ\'OCAR}

\address[]{\orgdiv{Methods and Outreach}, \orgname{Novo Nordisk Pharma}, \orgaddress{\state{Madrid}, \country{Spain}}}

\corres{Antonio Remiro-Az\'ocar, Methods and Outreach, Novo Nordisk Pharma, Madrid, Spain. \email{aazw@novonordisk.com}. Tel: (+34) 91 334 9800} 

\presentaddress{Antonio Remiro-Az\'ocar, Methods and Outreach, Novo Nordisk Pharma, Calle V\'ia de los Poblados 3, Madrid, 28033, Spain}

\abstract{In evidence synthesis, effect modifiers are typically described as variables that induce treatment effect heterogeneity at the individual level, through treatment-covariate interactions in an outcome model parametrized at such level. As such, effect modification is defined with respect to a conditional measure, but marginal effect estimates are required for population-level decisions in health technology assessment. For non-collapsible measures, purely prognostic variables that are not determinants of treatment response at the individual level may modify marginal effects, even where there is individual-level treatment effect homogeneity. With heterogeneity, marginal effects for measures that are not directly collapsible cannot be expressed in terms of marginal covariate moments, and generally depend on the joint distribution of conditional effect measure modifiers and purely prognostic variables. There are implications for recommended practices in evidence synthesis. Unadjusted anchored indirect comparisons can be biased in the absence of individual-level treatment effect heterogeneity, or when marginal covariate moments are balanced across studies. Covariate adjustment may be necessary to account for cross-study imbalances in joint covariate distributions involving purely prognostic variables. In the absence of individual patient data for the target, covariate adjustment approaches are inherently limited in their ability to remove bias for measures that are not directly collapsible. Directly collapsible measures would facilitate the transportability of marginal effects between studies by: (1) reducing dependence on model-based covariate adjustment where there is individual-level treatment effect homogeneity or marginal covariate moments are balanced; and (2) facilitating the selection of baseline covariates for adjustment where there is individual-level treatment effect heterogeneity.}


\maketitle

\renewcommand{\thefootnote}{\alph{footnote}}

\section{Introduction}\label{sec1}

Indirect treatment comparisons and network meta-analyses synthesize the results of multiple trials.\cite{bucher1997results, sutton2008use, dias2013evidence, lumley2002network, lu2004combination, jansen2011interpreting, jansen2014indirect} The trials often target patient populations with different distributions of baseline characteristics. These differences may give rise to heterogeneous treatment effect sizes across studies, which threaten the validity of indirect comparisons. Random effects models are normally used to capture heterogeneity.\cite{dersimonian1986meta, dersimonian2007random, stijnen2010random, bowden2011individual, ades2005interpretation} However, the typical approaches do not explain heterogeneity or explicitly produce estimates in any specific population.\cite{dahabreh2020towards, sobel2017causal, barker2022causally} This is troublesome, given the fundamental role of the \textit{population} in formulating research questions in evidence synthesis and in scoping decision problems in health technology assessment (HTA), e.g.~in frameworks such as PICO (population, intervention, comparator, outcome).\cite{schardt2007utilization, julian2022can}

Covariate-adjusted indirect treatment comparisons\cite{signorovitch2010comparative, phillippo2018methods, phillippo2016nice, remiro2022parametric, remiro2022two} and network meta-regression approaches\cite{phillippo2020multilevel, jansen2012meta, cooper2009addressing, dias2013evidencedos, dias2011nice} can be used to account for differences in baseline characteristics between trials. In doing so, the methodologies generate treatment effect estimates in specific study samples or populations. When indirect comparisons are \textit{anchored} by a common comparator in a connected network of evidence, a crucial assumption involves the \textit{constancy} or \textit{transportability} of relative treatment effects between studies.\cite{phillippo2018methods, phillippo2016nice} For this assumption to hold, current covariate adjustment guidance recommends accounting for all covariates that are \textit{effect measure modifiers} in the analysis.\cite{jansen2011interpreting,jansen2014indirect, phillippo2018methods, phillippo2016nice, jansen2012directed, hoaglin2011conducting} These are covariates that induce variation between studies in relative treatment effects, as measured on a specific scale, within a pairwise contrast of treatments. 

There are different types of relative treatment effects; these can be \textit{marginal} or \textit{conditional}.\cite{remiro2021conflating, phillippo2021target, remiro2022target} Marginal effects quantify how mean outcomes change at the population level. Alternatively, conditional effects contrast mean outcomes between specific patients or subgroups of patients, conditioning on covariate patterns. Some of the most widely used relative effect measures in evidence synthesis are \textit{non-collapsible}, e.g.~odds ratios\cite{greenland1987interpretation} and hazard ratios.\cite{aalen2015does} Marginal and conditional estimands are generally not equal for non-collapsible effect measures.\cite{greenland2011adjustments, kaufman2010marginalia} In fact, it is possible that marginal and conditional estimands have different signs, a phenomenon sometimes referred to as Simpson's paradox.\cite{hernan2011simpson} 

Other widely used relative effect measures in evidence synthesis are \textit{collapsible}, such as mean differences, risk differences and risk ratios. A collapsible measure is one for which the marginal measure can always be expressed as a weighted average of conditional measures.\cite{huitfeldt2019collapsibility} For \textit{directly collapsible} measures, the collapsibility weights are determined by the marginal distributions of the covariates that are conditioned on.\cite{colnet2023risk} Difference measures such as mean differences and risk differences are directly collapsible, whereas the risk ratio is not.\cite{huitfeldt2019collapsibility, colnet2023risk} 

Whether an effect measure is collapsible or not, and whether it is directly collapsible or not, has implications on the class of covariates that are effect measure modifiers on the marginal scale, and on the types of covariates that may compromise the constancy of relative effects for the marginal measure. This article seeks to bring attention to the following: 
\begin{itemize}
\item In the absence of treatment effect heterogeneity at the individual level, marginal effects for non-collapsible measures such as the (log) odds ratio generally depend on the distribution of \textit{purely prognostic} covariates that are not effect measure modifiers on the conditional scale; 
\item In the presence of treatment effect heterogeneity at the individual level, marginal effects for measures that are not directly collapsible, such as the (log) risk ratio and the (log) odds ratio, generally depend on the full joint distribution of purely prognostic covariates and of covariates that are effect measure modifiers on the conditional scale.
\end{itemize}
Crucially, for certain types of summary measures, the set of marginal and conditional effect measure modifiers may not coincide. Namely, covariates that do not modify the conditional effect measure at the individual or subgroup level may impact the marginal effect measure at the population level. This has major implications for recommended practices in evidence synthesis when the target estimand is a marginal effect.

There has been active discussion about the preferred estimand when estimating relative treatment effects in HTA, and whether this should be marginal or conditional.\cite{remiro2022target, russek2022discussion, schiel2022commentary, spieker2022comments, senn2022conditions, van2022estimands, remiro2022some} While conditional estimands are useful to answer clinically relevant questions at the individual or subgroup level, marginal estimands are considered more appropriate from the perspective of health policy stakeholders making treatment and reimbursement decisions at the population level.\cite{remiro2022target, russek2022discussion, schiel2022commentary, spieker2022comments, remiro2022some} This article assumes that the inferential target of primary interest for population-level decisions in HTA is a marginal treatment effect. 

This article also highlights the importance of the ``summary effect measure'' in formulating research questions in evidence synthesis and, more generally, in HTA. Within the regulatory environment, the summary effect measure is at the heart of the estimands framework, as described by the E9 (R1) Addendum issued by the International Council of Harmonisation (ICH).\cite{mehrotra2016seeking, akacha2017estimands, akacha2017estimandstwo} Nevertheless, it is not an explicit component of frameworks such as PICO and often a secondary consideration when postulating research questions in HTA.\cite{remiro2022some} Arguably, such questions are incomplete without reference to a summary effect measure. The assessment of effect modifier status is directly tied to the scale of the selected summary measure, whether it is marginal or conditional, collapsible or not, or directly collapsible or not. As such, the constancy or transportability of relative effects will depend on different types of covariates for different summary measures. 

The paper is structured as follows. Section \ref{sec2} outlines in detail key concepts underlying the article. These are: marginal versus conditional estimands, model-free versus model-based estimands, collapsibility and direct collapsibility, and the implications that such aspects have to external validity and transportability. 

Section \ref{sec3} presents a simulation study that illustrates: (1) how, in the absence of treatment effect heterogeneity at the individual level, marginal treatment effects for non-collapsible measures can differ across populations; and (2) how, in the presence of treatment effect heterogeneity at the individual level, marginal treatment effects for measures that are not directly collapsible can depend on the distribution of purely prognostic covariates; namely, on the joint multivariate distribution of purely prognostic covariates and conditional effect measure modifiers. 

The simulation study illustrates these concepts empirically in an anchored two-study scenario, within the specific context of indirect treatment comparisons with restricted access to subject-level data. The performance of covariate-adjusted indirect comparisons is compared with that of indirect comparisons that do not adjust for covariates. Different outcome types, outcome-generating models and summary effect measures are considered. Section \ref{sec4} describes the results of the simulation study. 

Section \ref{sec5} discusses implications for current guidance on evidence synthesis when the target estimand is a marginal treatment effect. For non-collapsible effect measures, unadjusted anchored indirect comparisons can be biased and the use of covariate adjustment warranted, even in the absence of individual-level treatment effect heterogeneity. In the presence of such heterogeneity, the dependence of effect measures that are not directly collapsible on the joint distribution of covariates, including those that are not effect measure modifiers on the conditional scale, raises questions about the use of unadjusted anchored indirect comparisons. It also raises questions about the extent of error of covariate-adjusted anchored indirect comparisons with limited individual patient data. Finally, we make some concluding remarks in Section \ref{sec6}.

\section{Key concepts}\label{sec2}

\subsection{Marginal and conditional treatment effect estimands}\label{subsec21}

Consider that an ideal randomized controlled trial (RCT) has been conducted in a sample of patients, assumed to be representative (i.e.~a random sample) of a larger target population, defined by the inclusion/exclusion criteria of the trial. For a given subject, let $T$ denote a binary indicator of treatment assignment, taking a value of one ($T=1$) if the subject is assigned to the active treatment under investigation, and a value of zero ($T=0$) if the subject is assigned to the control group, e.g.~placebo or standard of care. Let $Y$ denote a clinical outcome of interest, and let $X$ denote a pre-treatment baseline covariate, measured at randomization, that is prognostic of the outcome irrespective of the treatment assigned to the subject. 

Within the regulatory environment, the use of \textit{estimands} has been stimulated by the publication of the ICH E9 (R1) Addendum, recognized by agencies such as the Food and Drug Administration in the United States.\cite{mehrotra2016seeking, akacha2017estimands, akacha2017estimandstwo} According to the addendum, an estimand is a ``population-level summary'', which precisely describes the treatment effect that is targeted by the clinical trial. The estimand should align with the scientific question posed by the study investigators and the study objective. While the ICH E9 (R1) Addendum describes several strategies to account for intercurrent (post-randomization) events, estimands in this article will follow the ``treatment policy'' strategy, closely related to the intention-to-treat principle. As such, the occurrence of intercurrent events is not considered relevant when defining the estimands of interest. 

While the ICH E9 (R1) Addendum does not explicitly use the term ``causal'', its language is certainly aligned with ``counterfactual'' reasoning. Therefore, we shall adopt the potential outcomes framework to formally define
the marginal and conditional treatment effect estimands.\cite{rubin1974estimating} Let $Y^t$ denote the potential outcome that would have been observed for a subject assigned to intervention $T=t$, with $t \in \{0, 1\}$. Each subject in the trial has two potential outcomes, $Y^1$ and $Y^0$, corresponding to different ``parallel worlds''. Assuming no dropout or loss to follow up, one of the potential outcomes is observed in the study. The other is the that would hypothetically be realized under a different treatment condition than that actually assigned. 

The \textit{marginal} average treatment effect estimand for the trial is a contrast between the, possibly transformed, means of the potential outcome distributions:\cite{van2022estimands}
\begin{equation}
ATE = g\left(E\left(Y^1 \right )\right) - g\left(E\left(Y^0 \right)\right),
\label{additive}
\end{equation}
where $E(\cdot)$ represents an expectation taken over the distribution of potential outcomes for the trial, and $g(\cdot)$ is an appropriate ``link'' function, mapping the mean potential outcomes onto the plus/minus infinity range. Having assumed that the study sample is a random sample of its underlying target population, no distinction will be made between estimands at the sample level and at the population level.  

A \textit{conditional} estimand for the treatment effect at $X=x$ may also be of interest, particularly if the treatment effect is expected to differ by values of the covariate. On the aforementioned additive scale, the conditional average treatment effect estimand for the trial is defined as:\cite{van2022estimands}
\begin{equation}
CATE = g\left(E\left(Y^1 \mid X=x \right)\right) - g\left(E\left(Y^0 \mid X=x \right)\right).
\label{cond_additive}
\end{equation}
In this case, the covariate plays ``an explicit role in the definition of the treatment effect''.\cite{van2023use} While the marginal estimand is the average treatment effect had everyone in the trial been assigned the active treatment versus the control, the conditional estimand is the average treatment effect had a subset of patients in the trial, with the same covariate profile $X=x$, been assigned the active treatment versus the control.\cite{greenland1987interpretation, van2023use} 

Because the study is a perfectly-executed randomized trial, the marginal $ATE$ estimand can be identified from the observed data without further assumptions using the unadjusted estimator $g\left(E\left(Y \mid T=1 \right )\right) - g\left(E\left(Y \mid T=0 \right)\right)$; that is, a comparison of, potentially transformed, average outcomes between treatment arms.\cite{van2022estimands, van2023use} Such estimator is unbiased because, by virtue of randomization, both treatment arms are balanced in expectation with respect to measured and unmeasured prognostic factors. When $X$ is binary or categorical, the conditional $CATE$ estimand within covariate stratum $X=x$, assuming this is ``on the support'' of the population targeted by the trial, can be identified as $g\left(E\left(Y \mid T=1, X=x \right )\right) - g\left(E\left(Y \mid T=0, X=x \right)\right)$. Conversely, when $X$ is continuous, identification will almost invariably require additional statistical modeling assumptions, at the risk of model misspecification bias when the proposed model is incorrect. 

In Equation \ref{additive} and Equation \ref{cond_additive}, the treatment effect summary measure has been defined on an additive scale. This is advocated for, almost unquestionably, by researchers in the evidence synthesis literature.\cite{dias2013evidence,phillippo2018methods, phillippo2016nice, van2013evidence, caldwell2012selecting, dias2018network} Typically: 
\begin{itemize}
\item For continuous outcomes $Y^t \in \mathbb{R}$, $g(\cdot)$ is the identity link and the average treatment effect is constructed on the mean difference scale, such that the marginal estimand in terms of potential outcomes is $E(Y^1) - E(Y^0)$; 
\item For non-negative integer (i.e., discrete ``count'') outcomes $Y^t \in \mathbb{N}$, $g(\cdot)$ is the logarithmic link and the average treatment effect is constructed on the log risk ratio (log relative risk) scale, such that the marginal estimand is $\ln{E(Y^1)} - \ln{E(Y^0)}$;\footnote{Accounting for time, the average treatment effect would be constructed on the log rate ratio (log incidence rate) scale, such that the marginal estimand is $\ln E(Y^1(\tau)) - \ln E(Y^0(\tau))$, where $\tau$ is the follow-up time of the trial, i.e.,~the exposure or offset, and $Y^t(\tau)$ denotes the potential number of events experienced over time $\tau$ for a subject assigned treatment $t \in \{0, 1\}$.}
\item For binary outcomes $Y^t \in \{0, 1\}$, $g(\cdot)$ is the logit link and the average treatment effect is constructed on the log odds ratio scale, such that the marginal estimand is $\ln {[ E(Y^1)/(1-E(Y^1))] } -\ln {[E(Y^0 )/(1-E(Y^0))}]$.
\end{itemize}

The choice of such measurement scales is influenced by the nature of the outcome and by a ``model-based'' view of estimands, which may be at odds with the ICH E9 (R1) Addendum. Nevertheless, this article's findings for each summary measure do not specifically apply to a given outcome type. For instance, when the outcome is binary, findings about the mean difference are also applicable to the risk difference, and conclusions about the (log) risk ratio also hold. 

\subsection{Model-based 
estimands and collapsibility}\label{subsec22}

The definitions of the marginal and conditional treatment effect estimands in Equation \ref{additive} and Equation \ref{cond_additive} are model-free,\cite{van2023use, vansteelandt2022assumption} despite the use of a link function. Such estimands are not necessarily encoded by coefficients in a parametric or semi-parametric model, and their interpretation does not necessarily rely on statistical assumptions about correct model specification.\cite{van2023use, vansteelandt2022assumption} Nevertheless, the preference of the evidence synthesis literature for such additive summary measures -- the mean difference for continuous outcomes, the log risk ratio for count outcomes, and the log odds ratio for binary outcomes -- is implicitly affected by modeling preferences for each outcome type. 

The outcome models used in meta-analysis typically belong to the generalized linear model (GLM) family.\cite{dias2018network} In GLM theory, the identity, log and logit functions are canonical link functions, which require effects to be additive on a specific linear predictor scale; that is, the mean difference, log risk ratio or log odds ratio scale, respectively. Dias et al. state that ``it is important to use a scale in which effects are additive, as is required by the (generalized) linear model'', and that ``choice of scale can be guided by goodness of fit or by lower between-study heterogeneity''.\cite{dias2018network} Similarly, van Valkenhoef and Ades emphasize that choosing ``a scale of measurement is not a matter of selecting a ``summary statistic'' on the basis of ease of interpretation or convenience, but one of choosing the most appropriate statistical model for the data at hand''.\cite{van2013evidence}  

Admittedly, Dias et al. do acknowledge that ``quite distinct from choice of scale for modeling is the issue of how to report treatment effects'', and that investigators are ``free to derive treatment effects on other scales''.\cite{dias2013evidence, dias2018network} Nevertheless, as highlighted by Caldwell et al., while the ``choice of scale for analysis (...) should be kept distinct from the issue of which scale to use to report treatment effects'', in practice ``this is rarely carried out''.\cite{caldwell2012selecting} 

A common practice in evidence synthesis is to postulate a hypothetical outcome-generating model.\footnote{Arguably, there is a misalignment between certain practices in meta-analysis and the ICH E9 (R1) Addendum. Firstly, such addendum requires the estimand to be a summary measure that is defined without reference to a particular statistical model. Secondly, it requires one to specify the estimand, on the basis of research objectives, prior to selecting the statistical method for estimation. The addendum favors separating the definition of the estimand from the selection of the estimator, but current practices in evidence synthesis do not seem compatible with this sequential approach.} Let's assume that, by divine revelation, this is known to be a parametric GLM, specifying the conditional expectation of the potential outcome $Y^t$ under $T=t$ given $X$, at the individual level: 
\begin{equation}
E (Y^{t} \mid X) = g^{-1} \left( \beta_0 + \beta_X X + 
\beta_T t \right),  
\label{eqn1}
\end{equation}
where $g(\cdot)$ is a suitable invertible link function, $\beta_0 \in \mathbb{R}$ is an intercept term, and the model parameters $\beta_X, \beta_T \in \mathbb{R}$ are non-null coefficients quantifying conditional predictor-outcome associations. The model form in Equation \ref{eqn1} does not contain a product term -- that is, a statistical interaction -- between the baseline covariate and the intervention. Therefore, covariate $X$ is deemed to be \textit{purely prognostic},\cite{phillippo2018methods, phillippo2016nice, remiro2022parametric, remiro2021methods, liu2022correct} and there is said to be treatment effect \textit{homogeneity} at the individual level. 

Firstly, let the outcome-generating mechanism be a linear model, such that $g(\cdot)$ is the identity link in Equation \ref{eqn1}, as is typically postulated when the outcome is continuous. Under such model, the conditional average treatment effect estimand on the mean difference scale is:
\begin{align*}
CATE_{\textnormal{MD}} &= E\left(Y^1 \mid X=x \right) -E\left(Y^0 \mid X=x \right) \\
&=  
\beta_0 + \beta_X x + \beta_T - 
\left (
\beta_0 + \beta_X x
\right )
\\
&= \beta_T,
\end{align*}
The marginal average treatment effect estimand on the mean difference scale is: 
\begin{align*}
ATE_{\textnormal{MD}} 
&= E \left(Y^1  \right)  - 
E \left(Y^0 \right)  \\
&= E_X \left[ E \left(Y^1 \mid X \right) \right] -E_X \left [ E  \left(Y^0 \mid X \right) \right ] \\
&=  
E_X \left [\beta_0 + \beta_X X + \beta_T \right ] - E_X \left [\beta_0 + \beta_X X \right]\\
&= 
\beta_0 + \beta_XE_X [X] + \beta_T - 
\left (
\beta_0 + \beta_X E_X[X]  
\right ) \\
&= \beta_T,
\end{align*}
where $E_X[h(X)] = h\left[E_X(X)\right]$ for a linear function $h(\cdot)$. Therefore, the coefficient $\beta_T$ is interpretable both as a marginal and a conditional average treatment effect estimand, and the model-based conditional estimand coincides with the model-free definition of both the marginal and the conditional mean difference. Notably, under the specified outcome-generating model, the marginal estimand on the mean difference scale does not depend on the distribution of the purely prognostic covariate $X$. 

Next, let's assume that the outcome-generating mechanism is a log-linear model, e.g.~a Poisson model, such that $g(\cdot)$ is the logarithmic link function in Equation \ref{eqn1}, as is typically considered in the analysis of count outcomes. Under such model, the conditional average treatment effect estimand on the log risk ratio scale is:
\begin{align*}
CATE_{\textnormal{log RR}} &= \ln \left [ E\left(Y^1 \mid X=x \right) \right ] - \ln \left [ E\left(Y^0 \mid X=x \right) \right ]\\
&= \ln \left [ \exp \left (
\beta_0 + \beta_X x + \beta_T \right ) \right ]
- \ln \left [ \exp \left (
\beta_0 + \beta_X x \right ) \right ]
\\
&=  
\beta_0 + \beta_X x + \beta_T - \left (\beta_0 + \beta_X x \right )\\
&= \beta_T.
\end{align*}
The marginal average treatment effect estimand on the log risk ratio scale is:
\begin{align*}
ATE_{\textnormal{log RR}}
&= \ln \left [ E \left(Y^1  \right) \right ]  - 
\ln \left [ E \left(Y^0 \right) \right ]  \\
&= \ln \left \{ E_X \left[ E \left(Y^1 \mid X \right) \right] \right \} - \ln \left \{ E_X \left[ E \left(Y^0 \mid X \right) \right] \right \} \\
&=  
\ln \left \{
E_X \left [  \exp \left (  \beta_0 + \beta_X X + \beta_T \right ) \right ] 
\right \}
- 
\ln \left \{
E_X \left [\exp \left (  \beta_0 + \beta_X X \right ) \right]
\right \}
\\
&=  
\ln \left \{
E_X \left [  \exp \left ( \beta_0 + \beta_T
\right ) \exp \left (\beta_X X \right)  \right ] 
\right \}
- 
\ln \left \{
E_X \left [\exp \left (  \beta_0  \right) \exp \left (\beta_X X \right )  \right]
\right \}
\\
&=  \ln \left \{ \exp
\left ( \beta_0+\beta_T \right )
E_X \left [\exp\left (\beta_XX \right ) \right] \right \} - \ln \left \{ \exp \left (\beta_0 \right )E_X \left [\exp \left (\beta_XX \right ) \right ] \right \}  \\
&=  \ln \left \{  
\frac{
\exp
\left ( \beta_0 \right ) \exp \left (\beta_T \right )
E_X \left [\exp\left (\beta_XX \right ) \right]} { \exp \left (\beta_0 \right )E_X \left [\exp \left (\beta_XX \right ) \right ]} \right \}  
=
\ln \left [      
\exp \left (\beta_T \right) 
\right]
=
\beta_T
\end{align*}
where $E_X[d \exp(X)] = d  E_X[\exp(X)]$ for any constant $d \in \mathbb{R}$. Again, the coefficient $\beta_T$ can be interpreted both as a marginal and a conditional average treatment effect estimand,\cite{daniel2021making, vellaisamy2008collapsibility} and the model-based conditional estimand coincides with the model-free definition of both the marginal and the conditional log risk ratio. In addition, under the specified outcome-generating model, the marginal estimand on the log risk ratio scale does not depend on the distribution of the purely prognostic covariate $X$.

Finally, suppose that the outcome-generating mechanism  is a logistic model, such that the link function $g(\cdot)$ in Equation \ref{eqn1} is $\logit \left(p \right)=\ln \left [p/(1-p) \right ]$ for outcome probability $p \in [0, 1]$. In practice, this is a model commonly considered when the outcome is binary. Under the specified outcome-generating model, the conditional average treatment effect estimand on the log odds ratio scale is:
\begin{align}
\begin{split}
CATE_{\textnormal{log OR}} &= \logit \left [ E\left(Y^1 \mid X=x \right) \right ] - \logit \left [ E\left(Y^0 \mid X=x \right) \right ]\\ 
&= 
\logit \left [ \expit \left (
\beta_0 + \beta_X x + \beta_T \right ) \right ]
- \logit \left [ \expit \left (
\beta_0 + \beta_X x \right ) \right ]
\\
&=  
\beta_0 + \beta_X x + \beta_T - \left (\beta_0 + \beta_X x \right )\\
&= \beta_T, \label{eqn_cate_or}
\end{split}
\end{align}
where $\expit(\cdot)=\exp(\cdot)/[1+\exp(\cdot)]$ is the inverse logit function. We have observed that, for the linear and log-linear generative models, the parameter $\beta_T$ can be interpreted as a conditional as well as a marginal average treatment effect estimand. Conversely, the treatment coefficient of the logistic model corresponds to a conditional but not to a marginal estimand at the population level. This is a direct consequence of a mathematical property known as non-collapsibility.\cite{daniel2021making, freedman2008randomization, greenland1999confounding, morris2022planning} 

Non-collapsibility can be understood through the non-linearity of the characteristic collapsibility function (CCF),\cite{daniel2021making, neuhaus1993geometric} defined by Daniel et al.\cite{daniel2021making} as $f(\cdot) = g^{-1} \left [  g (\cdot) + m \right ]$, where $g(\cdot)$ is the link function of the GLM and $m$ is the conditional treatment-outcome association on the linear predictor scale, assumed constant across values of $X$ by the outcome-generating model in Equation \ref{eqn1}. The CCF maps $E \left (Y^0 \mid X \right )$ to $E \left (Y^1 \mid X \right )$. While the CCF is linear for the identity and logarithmic links, it is generally non-linear for the logit link.

Rearranging the expression in Equation \ref{eqn_cate_or}, we obtain:
\begin{align*}
\logit \left [ E\left(Y^1 \mid X \right) \right ] &= \logit \left [ E\left(Y^0 \mid X \right) \right ] + \beta_T,\\ 
E \left(Y^1 \mid X \right)  
&=
\expit \left \{
\logit \left[ E \left(Y^0 \mid X \right) \right ]
+ \beta_T 
\right \}.
\end{align*}
We let $f(p) = \expit \left [ \logit \left( p \right )+ \beta_T \right ]$ for all $p \in [0,1]$, such that $E \left (Y^1 \mid X \right )=f\left [ E \left (Y^0 \mid X \right ) \right]$. Following Daniel et al.\cite{daniel2021making} and Colnet et al.,\cite{colnet2023risk} we use Jensen's inequality such that, for $\beta_T > 0$:
\begin{align*}
E \left (Y^1 \right ) &=   E_X \left[ E \left(Y^1 \mid X \right) \right]  \\
&= 
E_X \left \{ f  \left [ E \left ( Y^0 \mid X \right )   \right ] \right \}
\\
&< f \left \{ E_X \left [  E  \left   ( Y^0 \mid X \right )   \right ] \right \}
\\
&= 
\expit \left \{
\logit \left[ E_X \left ( E \left(Y^0 \mid X \right) \right ) \right ]
+ \beta_T 
\right \} \\
&=
\expit \left \{
\logit \left[ E \left ( Y^0 \right )
\right ]
+ \beta_T 
\right \},
\end{align*}
because $f(\cdot)$ is concave for positive $\beta_T$. Conversely, for $\beta_T < 0$, $E\left (Y^1\right ) > \expit \left \{\logit \left[ E \left ( Y^0 \right ) \right ] + \beta_T  \right \}$, because $f(\cdot)$ is convex for negative $\beta_T$. As the logit link is a monotonous function, $\logit \left [ E \left (Y^1 \right ) \right ] < \logit \left [ E \left (Y^0 \right ) \right ] + \beta_T$ if $\beta_T > 0$, and $\logit \left [ E \left (Y^1 \right ) \right ] > \logit \left [ E \left (Y^0 \right ) \right ] + \beta_T$ if $\beta_T < 0$.\cite{colnet2023risk}

Accordingly, for $\beta_T > 0$, the marginal average treatment effect estimand on the log odds ratio scale is: 
\begin{align}
\begin{split}
ATE_{\textnormal{log OR}}
&= \logit \left [ E \left(Y^1  \right) \right ]  - 
\logit \left [ E \left(Y^0 \right) \right ]  \\
&< |\beta_T|
\label{not_pop_avg}
\end{split}
\end{align}
For $\beta_T < 0$, $ATE_{\textnormal{log OR}} > \beta_T$. In any case, having assumed that $\beta_X \neq 0$,\footnote{For $\beta_X = 0$, marginal and conditional estimands coincide for all effect measures because $E\left (Y^t \right ) = E \left (Y^t \mid X \right)$ for $t \in \left \{ 0, 1 \right \}$, such that there is collapsibility.} the conditional and marginal average treatment effect estimands are not equal on the log odds ratio scale for non-null $\beta_T$.\footnote{For $\beta_T = 0$, the marginal and conditional estimands are equal because the CCF is the identity function, therefore linear, irrespective of the link function.} In particular, $|ATE_{\textnormal{log OR}}| < |CATE_{\textnormal{log OR}}| = | \beta_T |$, which explains why the marginal estimand is closer to the null than the conditional estimand for the log odds ratio. 

Moreover, as we shall discuss in later sections of this article, the marginal (log) odds ratio generally depends on the full distribution of the purely prognostic covariate $X$, not just its mean, even where there is treatment effect homogeneity at the individual level. Importantly, one cannot generally reduce the expression of $ATE_{\textnormal{log OR}}$ to one that only involves $\beta_T$ or is analytically tractable over the covariate space, with numerical integration or simulation required to compute the marginal (log) odds ratio.\cite{austin2008performance}


In conclusion, insofar we have assumed that the conditional treatment effect on the linear predictor scale is the same for all individuals, regardless of the value of $X$. The mean difference and the (log) risk ratio are collapsible, which means that the population-level marginal estimand can be expressed as a (weighted) average of individual- or subgroup-level conditional estimands. As a result, when enforcing the constancy of the conditional estimand, the parameter $\beta_T$ has a population-level interpretation. Conversely, the (log) odds ratio is non-collapsible; almost invariably, the marginal measure cannot be expressed as a weighted average of conditional measures, even when the latter are constant.\cite{pang2016studying} As indicated by Equation \ref{not_pop_avg}, despite enforcing the homogeneity of the conditional (log) odds ratio across all subjects, the coefficient $\beta_T$ lacks any interpretation as a population-level average. It is certainly not a ``population-level summary'', using the language of the ICH E9 (R1) Addendum. 

Outside the GLM framework, this phenomenon has also been demonstrated for another non-collapsible effect measure, the (log) hazard ratio; see Daniel et al.\cite{daniel2021making}  and Section 3.2 of Martinussen and Vansteelandt.\cite{martinussen2013collapsibility} Namely, the marginal (log) hazard ratio generally depends on the distribution of purely prognostic covariates that do not ``interact'' with treatment, even under a ``proportional hazards'' generative model enforcing the homogeneity of the conditional (log) hazard ratio across all subjects. 

\subsection{Direct collapsibility}\label{subsec23}

In Section \ref{subsec22}, we have highlighted that a collapsible effect measure is one for which the marginal measure is always equal to a weighted average of conditional measures.\cite{huitfeldt2019collapsibility} When the effect measure is a mean difference, the collapsibility weights are determined by the marginal distributions of the covariates that are conditioned on. For instance, if the covariates are binary or categorical, the marginal mean difference is a weighted average of the subgroup-level conditional effects, with weights equal to the covariate probabilities, i.e., the proportion of subjects in each subgroup.\cite{colnet2023risk} As such, the mean difference is said to be \textit{directly collapsible}.\cite{colnet2023risk} 


In simpler terms, for difference measures such as the mean difference, the average over individual conditional differences is always equal to the difference in population-level marginal means, such that: 
\begin{equation*}
ATE_{\textnormal{MD}} = E \left (Y^1 \right )
- E \left (Y^0 \right ) 
= 
E_X \left [ E \left (Y^1 \mid X \right ) \right ] -
E_X \left [ E \left (Y^0 \mid X \right ) \right ] 
=
E_X \left [ E \left (Y^1 \mid X \right ) - E \left (Y^0 \mid X \right ) \right ].
\end{equation*}
Consider a study where 70\% of subjects, members of a specific subgroup, have a conditional mean difference of 0.2. The remaining 30\% of subjects, members of another subgroup, have a conditional mean difference of 0.6. The overall marginal mean difference across the study is weighted by the marginal covariate proportions/means, such that it is $0.7 \times 0.2 + 0.3 \times 0.6 = 0.32$.

In contrast, the risk ratio is not directly collapsible, such that the average over individual conditional ratios is not generally equal to the ratio of population-level marginal means.\cite{liu2022correct} That is:
\begin{equation*}
ATE_{\textnormal{RR}} = 
\frac{E \left (Y^1 \right )}
{E \left (Y^0 \right )} 
= 
\frac{E_X \left [ E \left (Y^1 \mid X \right ) \right ] }
{E_X \left [ E \left (Y^0 \mid X \right ) \right ] } 
\neq
E_X \left [ \frac{ E \left (Y^1 \mid X \right ) }{E \left (Y^0 \mid X \right )} \right ].
\end{equation*}
Colnet et al. provide a simple proof using Jensen's inequality.\cite{colnet2023risk} Similarly, for the log risk ratio: 
\begin{align*}
ATE_{\textnormal{log RR}}
&= 
\ln{ \left [E \left (Y^1 \right ) \right ]}
- \ln{ \left [E \left (Y^0 \right ) \right ] } \\
&= 
\ln{ \left \{
E_X \left [ E \left (Y^1 \mid X \right ) \right ] \right \} }
-
\ln {  \left \{ 
E_X \left [ E \left (Y^0 \mid X \right ) \right ] \right \}   } \\
&\neq 
E_X  \left \{
\ln{ 
\left [ E \left (Y^1 \mid X \right ) \right ] }
-
\ln { \left [ E \left (Y^0 \mid X \right ) \right ]} \right \}. 
\end{align*}
Collapsibility for the (log) risk ratio requires a different, more complex, set of weights than for the mean difference.\cite{huitfeldt2019collapsibility} As stated by Huitfeldt et al., the marginal risk ratio is ``generally not equal to a weighted average of the conditional (...) risk ratios, if the weights are determined by the marginal distribution of the covariates''.\cite{huitfeldt2019collapsibility} 

To illustrate how the (log) risk ratio is not directly collapsible, we move away from the ``treatment effect homogeneity'' scenario in Equation \ref{eqn1}, and assume that the postulated outcome-generating model has the following form:
\begin{equation}
E (Y^{t} \mid X) = g^{-1} \left( \beta_0 + \beta_X X + 
\beta_T t + \beta_{XT}X t \right).  
\label{eqn1_interaction}
\end{equation}
In this case, $X$ is a prognostic covariate, with the model parameter $\beta_X \in \mathbb{R}$ quantifying the conditional association in the control group between such covariate and the outcome. Assuming the coefficient $\beta_{XT} \in \mathbb{R}$ for the product term is non-null, there is a treatment-covariate interaction, and covariate $X$ also modifies the conditional effect of treatment on outcome on the linear predictor scale.\footnote{Note that, while the concepts of interaction, effect (measure) modification and effect heterogeneity are often used interchangeably in the biostatistics literature, they are not synonymous in the causal inference literature, where they may invoke slightly different mechanisms.\cite{vanderweele2009distinction, vanderweele2007four, hernan2020causal}} 

Because the conditional treatment effect depends on the level of $X$, the covariate is said to induce treatment effect \textit{heterogeneity} at the subject level, and is referred to as a (conditional) \textit{effect modifier}.\cite{vanderweele2009distinction, vanderweele2011interpretation, longford1999selection} We use the term \textit{effect measure modifier} because the presence of effect modification is contingent on the scale used to measure the relative treatment effect.\cite{brumback2008effect} Adopting terminology from the literature on biomarkers, $X$ is prognostic of outcome and also ``predictive'' of treatment response at the individual level.\cite{liu2022correct}\footnote{Some may find unsatisfactory the use of a parametric modeling framework to describe effect
measure modification. In the evidence synthesis literature, the link function of a parametric model typically influences the measurement scale on which effect modification is evaluated, such that effect modifier status is a function of the parameters of the hypothetical outcome-generating model.\cite{phillippo2018methods, phillippo2016nice, dias2013evidencedos, dias2011nice} The selected scale is, almost ubiquitously, the linear predictor scale used for parameter estimation,\cite{dias2013evidence, dias2013evidencedos,  dias2011nice} implicitly assuming that this coincides with the effect measure used to summarize the target estimand. The causal inference literature deems preferable a model-free definition of effect measure modification. This is typically based on counterfactuals contrasted at the individual or subgroup level on the scale of the target estimand.\cite{colnet2023risk, vanderweele2007four, hernan2020causal}} 

On the linear predictor scale and under the generative model in Equation \ref{eqn1_interaction}, the conditional treatment effect estimand of $T=1$ versus $T=0$, given baseline covariate $X=x$, is: 
\begin{align*}
CATE  &= 
g \left [
E\left(Y^1 \mid X=x\right) \right ] -
g \left [
E\left(Y^0 \mid X=x \right) \right ] \\
&=  
g \left [
g^{-1} \left (
\beta_0 + \beta_X x + \beta_T + \beta_{XT} x \right ) \right ]
-
g \left [
g^{-1} \left (
\beta_0 + \beta_X x
\right ) 
\right ] \\
&= 
\beta_0 + \beta_X x + \beta_T + \beta_{XT} x
- 
\left (\beta_0 + \beta_X x \right ) \\
&= \beta_T + \beta_{XT} x.
\end{align*}
There is no longer a single conditional estimand as this depends on the value of the covariate. For the linear generative model, the marginal average treatment effect estimand of $T=1$ versus $T=0$ on the mean difference scale is:
\begin{align*}
ATE_{\textnormal{MD}} 
&= E \left(Y^1  \right)  - 
E \left(Y^0 \right)  \\
&= E_{X} \left[ E \left(Y^1 \mid X \right) \right] -E_{X} \left [ E  \left(Y^0 \mid X \right) \right ] \\
&=  
E_{X} \left [\beta_0 + \beta_X X + \beta_T + \beta_{XT} X \right ] - E_{X} \left [\beta_0 + \beta_X X \right]\\
&= 
\beta_0 + \beta_XE_{X} [X] + \beta_T + \beta_{XT} E_{X}[X] - 
\left (
\beta_0 + \beta_X E_{X}[X]  
\right ) \\
&= \beta_T + \beta_{XT} E_{X}[X].
\end{align*}
Due to the presence of (conditional) effect measure modification by $X$, $\beta_T$ no longer has a marginal interpretation. Nevertheless, the expression of the marginal average treatment effect estimand for the mean difference can be reduced to one that only involves $\beta_T$, $\beta_{XT}$ and $E_{X}[X]$. Because the mean difference is directly collapsible, the marginal estimand can be expressed simply in terms of the model coefficients and the marginal covariate information, in this case the mean or proportion of $X$ in the study. 

Following Kiefer and Mayer,\cite{kiefer2019average} we shall illustrate by contradiction that, unlike the mean difference, the (log) risk ratio is not directly collapsible. Let's assume that the risk ratio is directly collapsible over $X$, such that: 
\begin{equation}
E_{X}\left[ \frac{E_{X}(Y^1 \mid X)}{E_{X}(Y^0 \mid X)} \right ] 
=
\frac{E_{X}\left [E_{X}(Y^1 \mid X) \right]}{E_{X} \left [E_{X}(Y^0 \mid  X) \right ]}. 
\label{ratio_dc}
\end{equation}
Substituting the parametric generative model for the conditional outcome expectation in Equation \ref{eqn1_interaction}, with $g(\cdot)= \ln(\cdot)$, into Equation \ref{ratio_dc}:
\begin{align}
E_{X}\left[ \frac{\exp \left (
\beta_0 + \beta_X X + 
\beta_T  + \beta_{XT}X \right )}{
\exp \left (\beta_0 + \beta_X X \right )} \right ] 
&=
\frac{E_{X} \left [\exp \left (
\beta_0 + \beta_X X + 
\beta_T  + \beta_{XT}X \right )
\right]}{E_{X} \left [\exp \left (\beta_0 + \beta_X X \right ) \right ]},\nonumber \\
\cancel{\frac{\exp \left ( \beta_0 + \beta_T \right )  }
{ \exp \left (\beta_0   \right )  }} \cdot
E_{X}\left[ \frac{ \cancel {\exp \left (
\beta_X X \right)} \exp \left  ( \beta_{XT}X \right )}{
\cancel{
\exp \left (\beta_X X \right ) }} \right ] 
&=
\cancel{\frac{\exp \left ( \beta_0 + \beta_T \right )  }
{ \exp \left (\beta_0   \right )  }}
\cdot 
\frac{ E_{X} \left[ \exp \left (
\beta_X X \right ) \exp \left ( \beta_{XT}X \right ) \right ]  }{
E_{X} \left [ \exp \left ( \beta_X X  \right ) \right ]}, \nonumber \\
E_{X}
\left [
\exp \left ( \beta_{XT}X \right ) 
\right ]
&= 
\frac{ E_{X} \left[ \exp \left (
\beta_X X \right ) \exp \left ( \beta_{XT}X \right ) \right ]  }{
E_{X} \left [ \exp \left ( \beta_X X  \right ) \right ]}, \nonumber \\
E_{X} \left [ \exp \left ( \beta_X X  \right ) \right ] \cdot
E_{X}
\left [
\exp \left ( \beta_{XT}X \right ) 
\right ]
&= 
E_{X} \left[ \exp \left (
\beta_X X \right ) \exp \left ( \beta_{XT}X \right ) \right ].\label{eqn7}
\end{align}
The covariance between two terms is equal to the expected value of their product minus the product of their expected values: 
\begin{equation}
\textnormal{Cov} 
\left [
\exp \left ( \beta_X X \right ), 
\exp \left (\beta_{XT} X \right )
\right ]
=
E_{X}
\left [
\exp \left ( \beta_X X \right )
\exp \left (\beta_{XT} X \right )
\right ]
-
E_{X}
\left [
\exp \left ( \beta_X X \right )
\right ]
\cdot
E_{X}
\left [
\exp \left (\beta_{XT} X \right )
\right ].
\label{eqn8}
\end{equation}
Equation \ref{eqn7} and Equation \ref{eqn8} imply that: 
\begin{equation}
\textnormal{Cov} 
\left [
\exp \left ( \beta_X X \right ), 
\exp \left (\beta_{XT} X \right )
\right ]
= 0.\label{eqn9}
\end{equation}
Nevertheless, Equation \ref{eqn9} only holds: (1) when $\beta_X=0$; or (2) when $\beta_{XT}=0$.\cite{kiefer2019average} 

That is, having assumed that the first condition does not apply, such that $X$ is prognostic of outcome in the control group, the risk ratio is only directly collapsible across $X$ if and only if there is no (conditional) effect measure modification by $X$. This would correspond to the log-linear generative model in Section \ref{subsec22}, where the marginal (log) risk ratio is equal to the average over conditional (log) risk ratios because the conditional effect measures are equal across covariate values, and the marginal (log) risk ratio is equal to all conditional (log) risk ratios. This is a  special case, only arising from enforcing treatment effect homogeneity at the individual level on the (log) risk ratio scale. 

Where there is treatment effect heterogeneity, under the outcome-generating model in Equation \ref{eqn1_interaction} with $g(\cdot) = \ln(\cdot)$, the marginal (log) risk ratio cannot simply be identified by $\beta_T$, $\beta_{XT}$ and $E_X (X)$. As we shall show empirically in a simulation study in this article, this is particularly important when considering multiple baseline covariates. In that case, in the presence of treatment effect heterogeneity at the individual level, the marginal (log) risk ratio does not only depend on summary moments for the marginal distribution of the (conditional) effect measure modifiers. It depends on the full joint covariate distribution (means, variances, covariance structure, distributional forms, etc.) of both: (1) covariates that are (conditional) effect measure modifiers, directly inducing subject-level treatment effect heterogeneity; and (2) purely prognostic covariates that do not directly induce such heterogeneity but are associated with the (conditional) effect measure modifiers.

\subsection{Implications for external validity and transportability}\label{subsec24}

Consider the ideal RCT described in Section \ref{subsec21}. Let's denote it the ``index'' trial or $S=1$. The marginal average treatment effect estimand, within index study $S=1$, is defined as:\cite{van2022estimands}
\begin{equation}
SATE = g\left(E\left(Y^1 \mid S=1 \right )\right) - g\left(E\left(Y^0 \mid S=1 \right)\right).
\label{additive_study}
\end{equation}
As described in Section \ref{subsec21}, the $SATE$ can be identified as a consequence of the study's high internal validity. Nevertheless, the target population that is of interest to researchers may differ from the study sample and from the underlying target population of $S=1$, in which case there may be limited \textit{external validity}.\cite{van2022estimands} More precisely, there may be limited \textit{transportability} if the target population that is of interest to researchers contains patients who are not eligible for enrollment in $S=1$, according to the selection criteria of the study.\cite{degtiar2023review} 

Establishing the external validity and transportability of study results is an essential part of HTA.\cite{schiel2022commentary} For instance, HTA agencies may demand extending the treatment effect estimates of a clinical trial beyond the study sample, to a broader ``real-world'' target population that is more relevant to their remit.\cite{happich2020reweighting} Another prevalent task involves transporting inferences from a randomized trial to an external study to perform an ``anchored indirect comparison'' between treatments that have not been compared in a head-to-head trial.\cite{phillippo2018methods} This is a scenario that is explored in the simulation study in Section \ref{sec3}. Regardless of the specific scenario, the external target shall be denoted $S=2$. 

In the aforementioned examples, the marginal average treatment effect estimand of interest, on the additive scale, would be that within the target $S=2$:\cite{van2022estimands} 
\begin{equation}
TATE = g\left(E\left(Y^1 \mid S=2 \right )\right) - g\left(E\left(Y^0 \mid S=2 \right)\right).
\label{additive_target}
\end{equation}
To transfer inferences from $S=1$ to $S=2$, the evidence synthesis literature typically invokes a \textit{constancy} or \textit{consistency} assumption.\cite{phillippo2018methods, phillippo2016nice, lu2006assessing} That is, the average treatment effect for active treatment versus control is the same in $S=1$ and in $S=2$. Admittedly, the literature does not often specify whether such treatment effect is marginal or conditional. Under the assumption that the target estimand is marginal, the constancy assumption holds if the following equality is met:
\begin{equation}
SATE = TATE, 
\label{constancy_assumption}
\end{equation}
where $SATE$ is defined as per Equation \ref{additive_study} and $TATE$ is defined as per Equation \ref{additive_target}. 

If covariate data that are representative of the external target are available, covariate adjustment techniques such as weighting, outcome model-based standardization or doubly-robust combinations of both, have been proposed to relax the constancy assumption.\cite{signorovitch2010comparative,phillippo2018methods, phillippo2016nice, remiro2022parametric,  remiro2022two} Because the trial comparing treatment $T=1$ and $T=0$ has not been performed in $S=2$, covariate adjustment becomes imperative to identify the $TATE$. Inevitably, the assumptions required to identify the $TATE$ from the observed data are not implied by randomization and are more stringent than those needed to identify the $SATE$. In particular, all effect measure modifiers, on the scale of interest, must be accounted for when transporting treatment effect estimates from the index study to the external target.

A central premise of this article, following Section \ref{subsec22} and Section \ref{subsec23}, is to show that, depending on the summary effect measure, different types of covariates determine whether the constancy assumption in Equation \ref{constancy_assumption} holds. More precisely, in the absence of individual-level treatment effect heterogeneity on the scale of the selected summary measure: 
\begin{itemize}
\item For collapsible effect measures such as the mean difference and the (log) risk ratio, the constancy assumption holds for the marginal measure. When the conditional measure is homogeneous across covariate values, the marginal measure does not depend on the distribution of purely prognostic covariates.
\item For non-collapsible effect measures such as the (log) odds ratio, the constancy assumption may not hold. Even if the conditional measure is constant across covariate values, the marginal measure generally depends on the distribution of purely prognostic covariates.  
\end{itemize}
Conversely, in the presence of treatment effect heterogeneity at the individual level:
\begin{itemize}
\item For directly collapsible effect measures such as the mean difference, marginal measures only depend on the distribution of (conditional) effect measure modifiers. The constancy assumption is only compromised by differences in the distribution of covariates belonging to such class. 
\item For effect measures that are not directly collapsible such as the (log) risk ratio and the (log) odds ratio, the marginal measure generally depends on the joint multivariate distribution of purely prognostic covariates and (conditional) effect measure modifiers. Consequently, the constancy assumption can be compromised by differences in such joint distribution. 
\end{itemize}
Interestingly, in the ``heterogeneity'' scenario and as illustrated empirically in Section \ref{subsec322}, the marginal (log) risk ratio, which is collapsible but not directly collapsible, does not seem to depend on the distribution of purely prognostic covariates when these are not associated with the (conditional) effect measure modifiers. Conversely, as a result of its non-collapsibility, the marginal (log) odds ratio does depend on the distribution of purely prognostic covariates, even if these are not associated with the (conditional) effect measure modifiers. Webster-Clark and Keil have recently arrived to similar conclusions.\cite{webster2023choice}

In summary, whether an effect measure is collapsible or not, and whether it is directly collapsible or not, has implications on the type of covariates that induce treatment effect heterogeneity on the marginal scale. Crucially, for non-collapsible effect measures such as the (log) odds ratio, purely prognostic covariates can act as \textit{marginal effect measure modifiers} at the population level, even if they are not (conditional) effect measure modifiers and in the absence of treatment effect heterogeneity at the individual level. 

This phenomenon can also occur for measures that are collapsible, but not directly collapsible, such as the (log) risk ratio. In the presence of treatment effect heterogeneity at the individual level, such measures will generally depend on the joint distribution of purely prognostic covariates and (conditional) effect measure modifiers. As shall be discussed in Section \ref{sec5}, in the specific context of indirect treatment comparisons with limited patient-level data, these findings have implications for recommended covariate adjustment practices when transporting inferences from $S=1$ to $S=2$. 

\section{Simulation study}\label{sec3} 

A simulation study is designed using the ADEMP (Aims, Data-generating mechanisms, Estimands, Methods, Performance measures) framework by Morris et al.\cite{morris2019using} The code required to conduct the simulation study is available online.\footnote{The files are available at 
\href{http://github.com/remiroazocar/conditional_marginal_effect_modifiers}{http://github.com/remiroazocar/conditional\_marginal\_effect\_modifiers}.} Simulations and analyses have been performed using \texttt{R} software version 4.1.1.\cite{team2013r}

\subsection{Aims}\label{subsec31}

The simulation study focuses on the specific setting of anchored indirect treatment comparisons in a two-study scenario with limited patient-level data. The empirical performance of covariate-adjusted indirect comparisons is compared with that of indirect comparisons that do not adjust for covariates. The following concepts are illustrated: 
\begin{enumerate}
\item In the absence of treatment effect heterogeneity at the subject level, marginal effects for non-collapsible measures such as the (log) odds ratio are not generally equal across populations with different distributions of purely prognostic covariates; 
\item In the presence of treatment effect heterogeneity at the subject level, marginal effects for measures that are not directly collapsible, such as the (log) risk ratio and the (log) odds ratio, are not generally equal across populations with different joint distributions of purely prognostic covariates and (conditional) effect measure modifiers. 
\end{enumerate}
In contrast, marginal effects for collapsible measures such as the mean difference and the (log) risk ratio are equal in the first setting, and marginal effects for directly collapsible measures such as the mean difference only depend on the distribution of (conditional) effect measure modifiers in the second setting. 

In the absence of treatment effect heterogeneity at the individual level, we shall observe that unadjusted anchored indirect comparisons can produce bias for non-collapsible measures when comparing marginal treatment effects across studies. In this case, the use of covariate-adjusted indirect comparisons that account for imbalances in purely prognostic covariates may be warranted. In the presence of treatment effect heterogeneity at the individual level, we shall observe that, when marginal covariate moments such as means and standard deviations are balanced across studies, unadjusted anchored indirect comparisons can still produce bias for measures that are not directly collapsible. In addition, while the use of covariate adjustment may be warranted with imbalanced marginal covariate summaries, bias can remain for measures that are not directly collapsible when failing to account for differences between the full joint covariate distributions, e.g.~correlations/covariances. 

\subsection{Data-generating mechanisms}\label{subsec32}

In each simulation, we generate data for two RCTs. Each RCT has two treatment arms and 5,000 participants, marginally randomized using a 1:1 allocation ratio. The studies are ideally-executed: there is perfect measurement and complete data. Randomization ensures that there is no structural confounding; in expectation, there is covariate balance between the treatment arms of each study. In addition, large sample sizes limit ``chance'' finite-sample imbalances within any particular simulated study. Large trials are simulated to show that the phenomena under investigation can afflict arbitrarily large datasets.  

Let $S=s$ denote a dichotomous study assignment indicator, such that $s \in \{1, 2\}$, and let $T=t$ denote a treatment assignment indicator, such that $t \in \{0, 1, 2\}$. Study $S=1$ compares active treatment $A$ ($T=1$) versus treatment $C$ ($T=0$), and study $S=2$ compares active treatment $B$ ($T=2$) versus treatment $C$. We seek an indirect comparison between $A$ and $B$, said to be anchored by common comparator $C$. We shall consider different data-generating mechanisms: one where there is treatment effect homogeneity at the individual level; another where there is treatment effect heterogeneity at the individual level and balance across studies in the marginal covariate distributions; and another whether there is treatment effect heterogeneity at the individual level and cross-study imbalances in the marginal covariate distributions. 

\subsubsection{Treatment effect homogeneity: imbalanced means and uncorrelated covariates}\label{subsec321}

For each of the subjects in the studies, three uncorrelated continuous baseline covariates $(X_1, X_2, X_3)$ are generated independently from normal marginal distributions with pre-specified means and standard deviations. Each baseline covariate is distributed differently across studies because there are imbalances in the marginal distribution means. For the $k$-th covariate, $X_k \sim \textnormal{Normal}(0, 1)$ in $S=1$, and $X_k \sim \textnormal{Normal}(-1.4, 1)$ in $S=2$. 

The following generative model for the conditional outcome expectation at the individual level, given treatment and covariates, is considered:
\begin{equation}
E \left (Y \mid X_1, X_2, X_3, T \right ) = g^{-1} \left( \beta_0 + \beta_1 X_1 + \beta_2X_2 + \beta_3 X_3 + \beta_T \mathbbm{1}\left (T=\textnormal{``active''} \right ) \right),
\label{ss_hom_eqn}
\end{equation}
where we select the intercept $\beta_0=-1$ and the main conditional covariate effects $\beta_1=\beta_2=\beta_3=1$ for both studies.\cite{austin2008performance} Three different outcome types, outcome-generating models and summary effect measures are examined: 
\begin{enumerate}
\item \textbf{Linear outcome model and mean difference}: a continuous outcome $Y \in \mathbb{R}$ is generated from the conditional expectation in Equation \ref{ss_hom_eqn}, with $g(\cdot)$ as the identity link, plus a residual error term from a standard (zero-mean, unit-variance) normal distribution. The summary measure for the average treatment effect is the mean difference.  
\item \textbf{Log-linear outcome model and log risk ratio}: a discrete count outcome $Y \in \mathbb{N}$ is generated from a Poisson distribution with outcome mean given by the conditional expectation in Equation \ref{ss_hom_eqn}, with $g(\cdot)$ as the logarithmic link. The summary measure for the average treatment effect is the log risk ratio. 
\item \textbf{Logistic outcome model and log odds ratio}: a binary outcome $Y \in \{0, 1\}$ is generated from a Bernoulli distribution with outcome probability given by the conditional expectation in Equation \ref{ss_hom_eqn}, with $g(\cdot)$ as the logit link. The summary measure for the average treatment effect is the log odds ratio.
\end{enumerate}
The outcome-generating model in Equation \ref{ss_hom_eqn} only contains main effects for the baseline covariates and lacks treatment-covariate product terms. As such, on the linear predictor scale, there is no treatment effect heterogeneity at the individual level or (conditional) effect measure modification by $X_1$, $X_2$ or $X_3$, which are said to be purely prognostic covariates. Consequently, the conditional treatment effect on the linear predictor scale for any of the active treatments versus the control is the same, $\beta_T$, for all subjects in a given study, regardless of covariate values. We have constructed the simulation setting so that the conditional treatment effects are equivalent for $A$ versus $C$ and $B$ versus $C$ in any study.

Following an iterative procedure by Austin and Stafford based on Monte Carlo integration,\cite{austin2008performance} we set the treatment coefficient $\beta_T=1.05$ to induce true marginal odds ratios of 2 and 2.45 (0.69 and 0.9 on the log odds ratio scale) in $S=1$ and $S=2$, respectively, for each active treatment versus control, in the third scenario with the logit link. 

In the first scenario with the identity link, due to the collapsibility of the mean difference and the absence of treatment effect heterogeneity at the individual level, the marginal mean difference for active treatment versus control in both $S=1$ and $S=2$ is equal to $\beta_T=1.05$. Similarly, in the second scenario with the log link, due to the collapsibility of the (log) risk ratio and the absence of treatment effect heterogeneity at the individual level, the marginal log risk ratio in both $S=1$ and $S=2$ is also equal to $\beta_T=1.05$. Notably, the marginal (log) odds ratio differs across studies with different distributions of purely prognostic covariates, but the marginal mean difference and the marginal (log) risk ratio do not, in the absence of (conditional) effect measure modification. 

\subsubsection{Treatment effect heterogeneity: balanced means and different correlation structures}\label{subsec322}

For each of the subjects in the studies, three continuous baseline covariates $(X_1, X_2, X_3)$ are generated from a multivariate normal distribution with pre-specified means, standard deviations and covariance matrix. This time, the marginal covariate distribution means are balanced across studies. For the $k$-th covariate, $X_k \sim \textnormal{Normal}(-1.4, 1)$ in $S=1$, and $X_k \sim \textnormal{Normal}(-1.4, 1)$ in $S=2$. Nevertheless, there are now differences between studies in the covariate correlation structures. In $S=1$, we set the pairwise linear correlation coefficients to $\textnormal{cor}\left (X_1, X_2 \right) = 0$, $\textnormal{cor}\left (X_1, X_3 \right ) =0$, and $\textnormal{cor} \left (X_2, X_3 \right ) =0$, such that the covariates are uncorrelated. In $S=2$, we set $\textnormal{cor} \left ( X_1, X_2 \right ) = 0$, $\textnormal{cor} \left ( X_1, X_3 \right ) = 0.4$, and $\textnormal{cor} \left ( X_2, X_3 \right ) = 0.4$, such that $X_1$ and $X_2$ are uncorrelated, but there is a moderate level of positive correlation between $X_3$ and each of the first two covariates. In summary, while there is balance between studies in the marginal distribution means and standard deviations, there are differences in the joint covariate distributions due to imbalances in the correlation coefficients. 

The following generative model for the conditional outcome expectation at the individual level, given treatment and covariates, is considered:
\begin{equation}
E \left (Y \mid X_1, X_2, X_3, T \right ) = g^{-1} \left( \beta_0 + \beta_1 X_1 + \beta_2X_2 + \beta_3 X_3 + \left ( \beta_T
+ \beta_{1T} X_1 \right ) \mathbbm{1}\left (T=\textnormal{``active''} \right ) \right),
\label{ss_het_eqn}
\end{equation}
We set $\beta_0=-1$, $\beta_1=\beta_2=\beta_3=1$, $\beta_{1T}=0.5$ and $\beta_T=1.05$ for both studies. While covariates $X_2$ and $X_3$ are purely prognostic, $X_1$ is prognostic of outcome in the control group, and also interacts with treatment. Therefore, it modifies the conditional effect of each active treatment versus control on the linear predictor scale, and induces treatment effect heterogeneity at the individual level on such scale.  

The three different outcome types, outcome-generating models and summary effect measures considered in Section \ref{subsec321} are examined, but with the conditional outcome expectation given by Equation \ref{ss_het_eqn}. For each scenario, true values of the marginal treatment effect in each study for active treatment versus control are determined by simulating a cohort of 50,000,000 subjects, a number sufficiently large to minimize sampling variability. Hypothetical individual-level outcomes under each treatment are generated for the cohort according to the true outcome-generating mechanism in Equation \ref{ss_het_eqn}. For each scenario, the true marginal mean difference, log risk ratio or log odds ratio is computed by averaging the simulated subject-level outcomes under each treatment, and contrasting the marginal outcome expectations on the corresponding linear predictor scale. 

In the first scenario, with $g(\cdot)$ as the identity link, the true marginal mean difference for active treatment versus control is 1.05 in both $S=1$ and $S=2$. In the second scenario, with $g(\cdot)$ as the log link, the true marginal log risk ratio is 0.97 and 1.18 in $S=1$ and $S=2$, respectively. In the third scenario, with $g(\cdot)$ as the logit link, the true marginal log odds ratio is 0.67 and 0.6 in $S=1$ and $S=2$, respectively. 

A simulation-based approach is necessary to compute the true marginal estimands in the second and third scenarios because the (log) risk ratio and (log) odds ratio are not directly collapsible. Such approach is not necessary in the first scenario because, due to its direct collapsibility, the true marginal mean difference can be expressed as a weighted average of the true conditional mean differences, with weights determined by the marginal covariate means. In this case, the true marginal mean difference in each study only depends on coefficients of the outcome model and the mean of the (conditional) effect measure modifier: $\beta_T + \beta_{1T} \times E (X_1 ) = 1.05 + 0.5 \times (- 1.4) = 0.35$, in both studies.  

Notably, in the presence of treatment effect heterogeneity at the individual level, the marginal (log) risk ratio and marginal (log) odds ratio differ across studies with identical marginal covariate means and standard deviations, due to differences in the covariate correlation coefficients. We make note of an important corollary. The marginal (log) risk ratio, which is collapsible but not directly collapsible, does not seem to depend on the distribution of purely prognostic covariates when these are uncorrelated with the (conditional) effect measure modifiers. For instance, in the second scenario with $g(\cdot)$ as the log link, consider setting $\textnormal{cor} \left ( X_1, X_2 \right ) = 0$, $\textnormal{cor} \left ( X_1, X_3 \right ) = 0$, and $\textnormal{cor} \left ( X_2, X_3 \right ) = 0.4$, such that only the purely prognostic covariates are correlated, and keeping the marginal covariate distributions unchanged. The true marginal log risk ratio is identical to that in $S=1$ with uncorrelated covariates (0.97). Conversely, the marginal (log) odds ratio is expected to depend on the distribution of purely prognostic covariates, even if these are not associated with the (conditional) effect measure modifiers.

\subsubsection{Treatment effect heterogeneity: imbalanced means and different correlation structures}\label{subsec323}

This setting is identical to that outlined in Section \ref{subsec322}, but with imbalances across studies in the marginal covariate distribution means. This time, for the $k$-th covariate, $X_k \sim \textnormal{Normal}(0, 1)$ in $S=1$, and $X_k \sim \textnormal{Normal}(-1.4, 1)$ in $S=2$. As per Section \ref{subsec322}, there are also differences between studies in the covariate correlation structures. In $S=1$, we set the pairwise linear correlation coefficients to $\textnormal{cor}\left (X_1, X_2 \right) = 0$, $\textnormal{cor}\left (X_1, X_3 \right ) =0$, and $\textnormal{cor} \left (X_2, X_3 \right ) =0$. In $S=2$, we set $\textnormal{cor} \left ( X_1, X_2 \right ) = 0$, $\textnormal{cor} \left ( X_1, X_3 \right ) = 0.4$, and $\textnormal{cor} \left ( X_2, X_3 \right ) = 0.4$.  In summary, the joint distribution of covariates is different across studies because there are imbalances in the marginal distribution means and in the correlation coefficients.

We use the generative model for the conditional outcome expectation at the individual level in Equation \ref{ss_het_eqn}. As per Section \ref{subsec322}, we set $\beta_0=-1$, $\beta_1=\beta_2=\beta_3=1$, $\beta_{1T}=0.5$ and $\beta_T=1.05$ for both studies, such that $X_1$ is a (conditional) effect measure modifier and prognostic of outcome in the control group, and $X_2$ and $X_3$ are purely prognostic covariates. The three different outcome types, outcome-generating models and summary effect measures considered in Section \ref{subsec321} and Section \ref{subsec322} are examined, with the conditional outcome expectation given by Equation \ref{ss_het_eqn}. 

True values of the marginal treatment effect in each study for active treatment versus control on the linear predictor scale are determined using the simulation-based approach outlined in Section \ref{subsec322}. In the first scenario, with $g(\cdot)$ as the identity link, the true marginal mean difference for active treatment versus control is 1.05 and 0.35 in $S=1$ and $S=2$, respectively. In the second scenario, with $g(\cdot)$ as the log link, the true marginal log risk ratio is 1.68 and 1.18 in $S=1$ and $S=2$, respectively. In the third scenario, with $g(\cdot)$ as the logit link, the true marginal log odds ratio is 0.69 and 0.6 in $S=1$ and $S=2$, respectively. 

\subsection{Estimands}\label{subsec33}

The target estimand is the true marginal treatment effect for $A$ versus $B$ in $S=2$, which is a composite of that for $A$ versus $C$ and that for $B$ versus $C$. The true marginal effect for active treatment versus control -- that is, $A$ versus $C$ or $B$ versus $C$ -- may vary across the settings of the simulation study. Nevertheless, in every simulation scenario, the true marginal effect in $S=2$ for $A$ versus $C$ is equal to the true marginal effect in $S=2$ for $B$ versus $C$. Because the summary effect measure is on the additive linear predictor scale -- either the mean difference, log risk ratio or log odds ratio scale -- the marginal effects for $A$ versus $C$ and for $B$ versus $C$ cancel out so that the true marginal effect for $A$ versus $B$ in $S=2$ is zero. 

\subsection{Methods}\label{subsec34}

The methods under evaluation operate in the following two-study scenario, common in HTA submissions.\cite{phillippo2016nice} The manufacturer submitting evidence for reimbursement has individual patient data from its own index study $S=1$, comparing the efficacy of a novel treatment $A$ versus control $C$. Conversely, the manufacturer only has access to aggregate-level summary data from the target study $S=2$, which has been conducted by an external party and compares a competitor treatment $B$ to control $C$. 

In practice, subject-level data for $S=2$ are unavailable due to privacy and confidentiality concerns. Only marginal summary moments are available for the covariates, e.g.~proportions for binary or categorical covariates and means with standard deviations for continuous covariates, sourced from a published table of clinical and demographic baseline characteristics. While cross-tabulations of discrete covariates are sometimes available in scientific manuscripts, information on the full joint covariate distribution in $S=2$, e.g.~distributional forms and correlation structure, is unlikely to be reported. To reflect the situation typically encountered by analysts, the individual-level covariates generated for $S=2$ are summarized as means with standard deviations, as would be available in the clinical trial publication. 

In anchored indirect comparisons, the marginal treatment effect for $A$ versus $B$ is estimated on the additive linear predictor scale -- mean difference, log risk ratio or log odds ratio scale -- as: 
\begin{equation}
\hat{\Delta}_{12} = \hat{\Delta}_{10} - \hat{\Delta}_{20},
\label{eqn5}
\end{equation}
where $\hat{\Delta}_{tt'}$ is an estimate of the marginal effect $\Delta_{tt'}$ for treatment $t$ versus $t'$. The following anchored indirect comparison methods will be compared: (1) matching-adjusted indirect comparison (MAIC); (2) parametric G-computation; and (3) the Bucher method. The first two use covariate adjustment to project inferences for $\Delta_{10}$ from $S=1$ to $S=2$, thereby performing the indirect comparison in $S=2$. MAIC is weighting-based and parametric G-computation is an outcome modeling-based approach. The Bucher method is the standard unadjusted anchored indirect comparison;\cite{bucher1997results, sutton2008use, dias2013evidence, lumley2002network, lu2004combination, jansen2011interpreting} it does not adjust for covariate differences in attempting to transport inferences between studies. An estimate of $\Delta_{10}$ is produced in $S=1$, using exclusively data of said trial. 

All three methods perform identical unadjusted analyses to estimate the marginal treatment effect for $B$ versus $C$, and all methods combine the relative effect estimates for $A$ versus $C$ and $B$ versus $C$ in the same manner. We shall outline these aspects first, prior to describing differences between the methods in the estimation of the marginal treatment effect for $A$ versus $C$. 

For all methods, the marginal treatment effect for $B$ versus $C$ in $S=2$, on the linear predictor scale, is estimated by fitting a simple generalized linear regression of outcome on treatment -- depending on the simulation scenario, either a normal linear regression, a Poisson regression or a logistic regression -- to the study's subject-level data. While such data are unavailable to the analyst, the point estimate of the treatment effect and its standard error would be reported in the clinical trial publication. Alternatively, these would be readily calculated from published summary tables. 

As a consequence of randomization, the unadjusted estimator is unbiased for the marginal effect estimand $\Delta_{20}$ in $S=2$. For all methods, relative effect estimates for $A$ versus $C$ and for $B$ versus $C$ are combined by plugging $\hat{\Delta}_{10}$ and $\hat{\Delta}_{20}$ into Equation \ref{eqn5}. Assuming statistical independence, point estimates of their variances are summed to estimate the variance of the marginal effect for $A$ versus $B$.\cite{bucher1997results, lumley2002network, lu2004combination, song2003validity} Wald-type 95\% confidence intervals are estimated using normal distributions. 

Next, we describe how the different anchored indirect comparison methods produce an estimate $\hat{\Delta}_{10}$ of the marginal treatment effect for $A$ versus $C$. Differences in statistical performance between the methods will arise from the estimation of such effect. 

\subsubsection{Matching-adjusted indirect comparison}\label{subsec341}

MAIC uses the method of moments (entropy balancing) approach originally proposed by Signorovitch et al.\cite{signorovitch2010comparative} to estimate the weights, as implemented by Remiro-Az\'ocar et al.\cite{remiro2022parametric} Covariate balance is viewed as a convex optimization problem. The BFGS algorithm is used to minimize the corresponding objective function. The estimated weights enforce exact balance between marginal moments of the weighted patient-level covariates for $S=1$ and those reported for $S=2$. 

We balance the sample means of the three baseline covariates across studies, for the active treatment and control arms combined. We do not balance the sample standard deviations because: (1) on expectation, these are already equal across studies; (2) to avoid unnecessary reductions in effective sample size and precision; and (3) to ensure that a solution to the convex optimization problem can be found. We only attempt to balance the marginal covariate distributions, not the joint covariate distributions, as correlation data are not typically published for $S=2$. As such, the covariate correlations and joint covariate distribution of $S=2$ are assumed to be equal to those of the weighted $S=1$ covariate data. 

The estimated weights are inputted to a weighted univariable generalized linear regression of outcome on treatment, fitted to the $S=1$ patient-level data. The regression is either a normal linear regression, a Poisson regression or a logistic regression, depending on the simulation study scenario. The treatment coefficient of the regression provides a point estimate of the marginal treatment effect for $A$ versus $C$ on the linear predictor scale. 

For variance estimation, we use the ordinary non-parametric bootstrap with replacement, with 500 resamples. This accounts for the correlation induced by weighting the $S=1$ observations and for the uncertainty in the weight estimation procedure. Both the weight estimation procedure and the estimation of the weighted generalized linear regression are included in each bootstrap iteration. The average marginal treatment effect for $A$ versus $C$ in $S=2$ is computed as the mean across the bootstrap resamples. Its standard error is the standard deviation across the resamples. 

\subsubsection{Parametric G-computation}\label{subsec342}

We use the maximum-likelihood version of parametric G-computation implemented by Remiro-Az\'ocar et al.\cite{remiro2022parametric} The performance of this method is expected to be competitive because the outcome model will be correctly specified. Parametric G-computation consists of several steps. 

\textbf{Covariate simulation}. As subject-level data are unavailable for $S=2$, the joint covariate distribution of the study is emulated based on its published summary statistics and on assumptions about the correlation structure and marginal distribution forms. Because correlations are not reported for $S=2$, its pairwise linear correlations are assumed to match those observed in the $S=1$ subject-level data, as has been recommended in the literature.\cite{remiro2022parametric, phillippo2020multilevel, phillippo2020assessing, ishak2015simulation} As the forms of the marginal covariate distributions in $S=2$ are unknown, these are typically selected on the basis of theoretical properties and the forms observed in $S=1$.\cite{remiro2022parametric, phillippo2020multilevel, phillippo2020assessing} We assume these to be normally-distributed. 1,000 individual-level covariate profiles are simulated from a multivariate Gaussian copula with normal marginal distributions, using the $S=2$ means and standard deviations, and the $S=1$ matrix of pairwise linear correlations.\cite{remiro2022parametric, phillippo2020multilevel, phillippo2020assessing}

\textbf{Model-fitting}. A multivariable generalized linear regression of outcome on treatment and baseline covariates is fitted to the $S=1$ subject-level data using maximum-likelihood estimation. The covariate-adjusted regression is correctly specified. Depending on the simulation scenario, it is either a normal linear regression, a Poisson regression or a logistic regression.

\textbf{Outcome prediction}. The fitted model is applied to all $S=2$ covariate profiles, to generate two counterfactual predictions of the conditional outcome expectation on the natural scale for each simulated subject. Treatment is set to $A$ or $C$ by manipulation: fixing the covariates at their simulated values, one prediction is under treatment $A$ and the other is under treatment $C$. 

\textbf{Average and contrast}. The two sets of predicted outcomes are averaged over the simulated covariate profiles to obtain estimates of the marginal outcome expectation under each treatment, on the natural scale. The averages are converted to the linear predictor scale and contrasted to obtain a point estimate of the marginal treatment effect for $A$ versus $C$ in $S=2$. 

For variance estimation, the $S=1$ subject-level data are resampled using the ordinary non-parametric bootstrap with replacement, with 500 resamples. Namely, it is the ``model-fitting'', ``outcome prediction'' and ``average and contrast'' steps that are iterated. The average marginal treatment effect for $A$ versus $C$ in $S=2$ is estimated as the mean across the bootstrap resamples. Its standard error is calculated as the standard deviation across the resamples.

\subsubsection{Bucher method}\label{subsec343}

The Bucher method is the standard unadjusted anchored indirect comparison, which does not account for covariate imbalances between studies.\cite{bucher1997results} This approach is usually deemed adequate in the absence of individual-level treatment effect heterogeneity or treatment-covariate interactions, or when (conditional) effect measure modifiers are equidistributed across studies.\cite{jansen2011interpreting, jansen2014indirect, phillippo2018methods, phillippo2016nice, cooper2009addressing, jansen2012directed, hoaglin2011conducting, remiro2021methods, remiro2022effect}

A simple generalized linear regression of outcome on treatment is fitted to the $S=1$ subject-level data. Depending on the simulation scenario, this is either a normal linear regression, a Poisson regression or a logistic regression. The model's treatment coefficient and nominal standard error give a point estimate of the marginal treatment effect for $A$ versus $C$ and its standard error, respectively, on the linear predictor scale.

\subsection{Performance measures}\label{subsec35}

For each simulation scenario, 500 datasets are simulated according to the data-generating mechanisms in Section \ref{subsec32}. Methodologies are assessed according to the following frequentist characteristics: (1) bias; (2) efficiency; and (3) coverage of interval estimates.\cite{morris2019using} The selected performance metrics specifically evaluate these criteria. We track: (1) bias; (2) mean square error (MSE), used to quantify efficiency; and (3) empirical coverage rate of the 95\% interval estimates, as defined by Morris.\cite{morris2019using} To characterize the simulation uncertainty, Monte Carlo standard errors over the simulation runs will be reported for each performance measure.\cite{morris2019using} Given the large number of subjects per simulated study, sampling variability should be small and 500 data replicates are expected to yield adequate simulation uncertainty. 

\section{Results}\label{sec4}

The results of the simulation study for the linear outcome model and the mean difference are summarized in Figure \ref{fig1}. Those for the log-linear outcome model and the log risk ratio are displayed in Figure \ref{fig2}, and those for the logistic outcome model and the log odds ratio are shown in Figure \ref{fig3}. In each of these figures, a ridgeline plot to the left visualizes the spread of point estimates over the 500 simulation replicates. To the right, a table reporting numeric values for the performance measures of each method is displayed, with MCSEs in parentheses alongside each performance measure.

\subsection{Linear outcome model and mean difference}\label{subsec41}
 
\begin{figure}[!htb]
\center{\includegraphics[width=\textwidth]{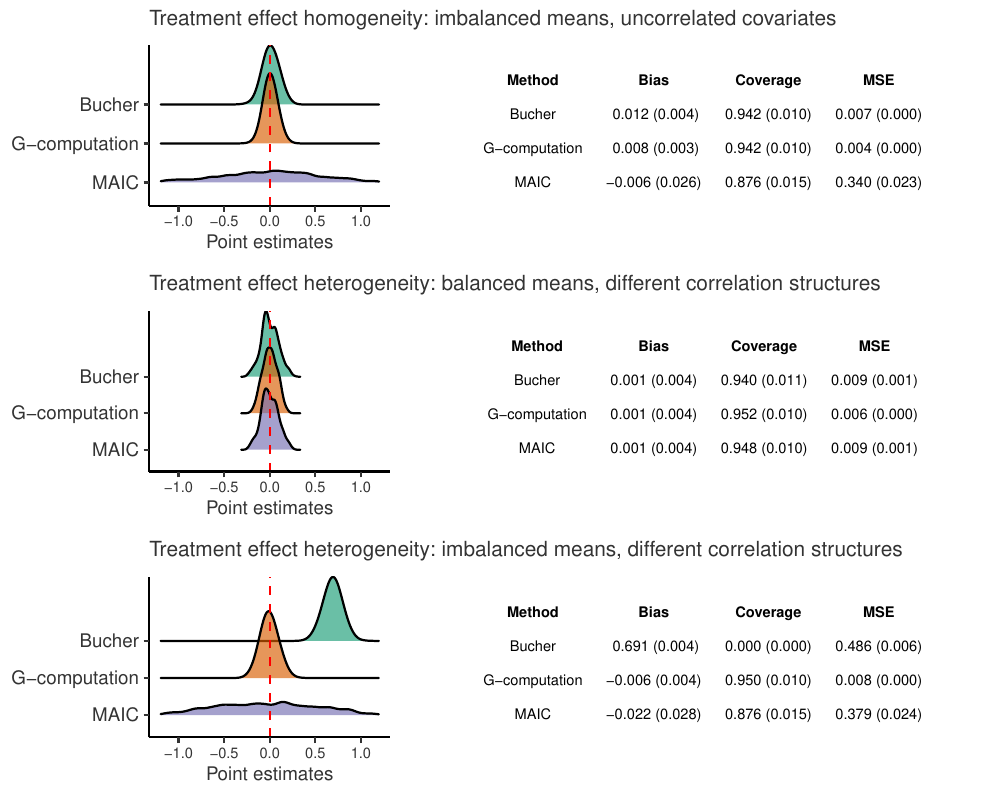}}
\caption[]
{\textbf{Linear outcome model and mean difference as the summary effect measure}. Point estimates of the marginal mean difference for $A$ versus $B$, and performance metrics with MCSEs for all methods across different settings.}
\label{fig1}
\end{figure}

For the linear outcome model with the mean difference as the summary measure (Figure \ref{fig1}), the performance of the Bucher method is competitive where there is treatment effect homogeneity at the individual level. In this setting, the method exhibits very little bias (0.012) and achieves an appropriate empirical coverage rate (0.942), within Monte Carlo error of the nominal 0.95 value. So is the case where there is treatment effect heterogeneity at the individual level and covariate means are balanced across studies. Here, the Bucher method is virtually unbiased (0.001) and exhibits an adequate empirical coverage rate (0.940), also within Monte Carlo error of the nominal value. Conversely, in the presence of treatment effect heterogeneity at the individual level, but with imbalanced covariate means, the performance of the Bucher method is deficient. There is substantial bias (0.691) and an extreme degree of undercoverage, as the 95\% confidence interval estimates do not contain the true marginal mean difference in any of the simulation replicates. 

Parametric G-computation offers negligible bias (0.008, 0.001 and -0.006) and valid confidence interval estimates (empirical coverage rates of 0.942, 0.952 and 0.950) in all three settings. In the two ``treatment effect heterogeneity'' settings, empirical coverage rates are virtually equal to the desired nominal value. MAIC produces negligible bias in all three settings (-0.006, 0.001 and -0.022), within Monte Carlo error of the true marginal mean difference. In the setting with balanced covariate means, MAIC provides excellent coverage (empirical coverage rate of 0.948). With imbalanced covariate means, MAIC exhibits undercoverage, with a coverage rate of 0.876 in both settings. Moreover, as a result of poor covariate overlap in these settings, the method is imprecise with large reductions in effective sample size after weighting. Despite the marked drop in precision, MAIC is still more efficient than the Bucher method in the ``heterogeneity'' setting with imbalanced means, due to the bias of the latter. 

The performance of parametric G-computation is comparable to that of the Bucher method in the settings with treatment effect homogeneity at the individual level, and with heterogeneity but balanced covariate means. Parametric G-computation even achieves a minor improvement in precision and efficiency in these settings. Nevertheless, these performance gains have required the correct specification of a parametric outcome model, a step that can be cumbersome in practice. Model misspecification would likely have implications in terms of bias, but the implications in the corresponding settings are unclear.

Generally, we can conclude that, in the absence of individual-level treatment effect heterogeneity or when influential marginal moments are balanced across studies, covariate adjustment is not necessarily warranted. Conversely, it can markedly improve performance with respect to the unadjusted approach where there is treatment effect heterogeneity at the individual level and marginal covariate moments are imbalanced across studies.  

\subsection{Log-linear outcome model and log risk ratio}\label{subsec42}

\begin{figure}[!htb]
\center{\includegraphics[width=\textwidth]{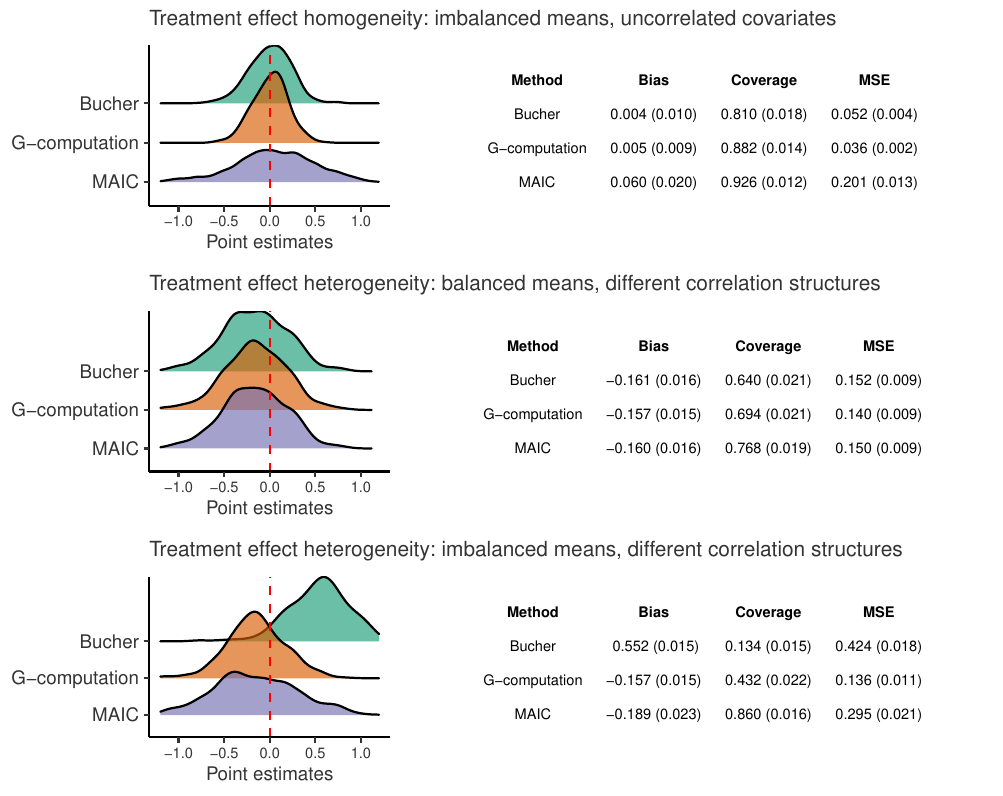}}
\caption[]
{
\textbf{Log-linear outcome model and log risk ratio as the summary effect measure}. Point estimates of the marginal log risk ratio for $A$ versus $B$, and performance metrics with MCSEs for all methods across different settings.
}
\label{fig2}
\end{figure}

For the log-linear outcome model with the log risk ratio as the summary measure (Figure \ref{fig2}), the Bucher method and parametric G-computation are virtually unbiased (0.005 and 0.004, respectively) where there is treatment effect homogeneity at the individual level. In this setting, MAIC exhibits some bias (0.060) but its empirical coverage rate (0.926) falls closer to the nominal 0.95 value than that of the other two methods, which display undercoverage (empirical coverage rates of 0.810 and 0.882 for the Bucher method and parametric G-computation, respectively). 

Where there is treatment effect heterogeneity at the individual level and correlations are unequal between studies, all approaches are biased and produce undercoverage. Importantly, even where the covariate means are perfectly balanced across studies, the Bucher method is biased (-0.161), as are parametric G-computation (-0.157) and MAIC (-0.160). MAIC does show improved coverage (0.768) with respect to the Bucher method (0.640) and parametric G-computation (0.694), which produce more discernible undercoverage.

\clearpage

Notably, the covariate adjustment approaches remain biased when accounting for all covariate mean imbalances in the third setting (bias of -0.157 for parametric G-computation and -0.189 for MAIC). There is also undercoverage, this being particularly troublesome for parametric G-computation (empirical coverage rate of 0.432) compared to MAIC (0.860). In any case, covariate adjustment does improve performance with respect to the Bucher method, which is liable to sizeable bias (0.552) and very poor coverage, with only 13.4\% of the 95\% confidence interval estimates covering the true marginal log risk ratio.  

Low empirical coverage rates may arise as a result of bias and/or overprecise standard errors. MAIC does not reduce bias compared to G-computation in any of the simulation settings. Nevertheless, it displays markedly improved coverage. With limited overlap, MAIC does not extrapolate, and its standard errors and interval estimates provide a more ``honest'' characterization of uncertainty. Parametric G-computation relies on modeling assumptions to extrapolate beyond the $S=1$ covariate space. Its estimated standard errors are overly precise and its confidence intervals are artificially narrow, even when taking into account the extent of model-based extrapolation. This warrants further investigation. Parametric G-computation's relatively high precision and efficiency, in terms of MSE, should not necessarily be viewed as a positive feature. Similarly, the imprecision of MAIC is not inherently undesirable, but rather an explicit manifestation of the high estimation uncertainty, particularly in the poor overlap settings with imbalanced covariate means. 

\clearpage

Undercoverage is most problematic for the Bucher method, which ignores any covariate differences between studies. Seemingly, empirical coverage rates can be degraded, even where existing covariate differences do not induce bias. An example is the setting with treatment effect homogeneity at the individual level. While bias is negligible and there is no bias-induced undercoverage, coverage is poor, likely due to variance underestimation. 

Generally, we can conclude that the unadjusted approach is inappropriate in the presence of treatment effect heterogeneity at the individual level. This is the case, even if marginal covariate moments are perfectly balanced across studies. Moreover, the unadjusted approach is systematically overprecise, across all settings. 

Covariate adjustment seems warranted where there are imbalances in marginal covariate moments. Nevertheless, questions are raised about only accounting for differences in marginal moments where there are differences across studies in correlation structures. To improve performance, accounting for differences in correlations -- more generally, in the full joint covariate distributions -- appears necessary. The over-precision of parametric G-computation is substantial, even when taking into account the extent of model-based extrapolation, and warrants further investigation. 

\clearpage

\subsection{Logistic outcome model and log odds ratio}\label{subsec43}

\begin{figure}[!htb]
\center{\includegraphics[width=\textwidth]{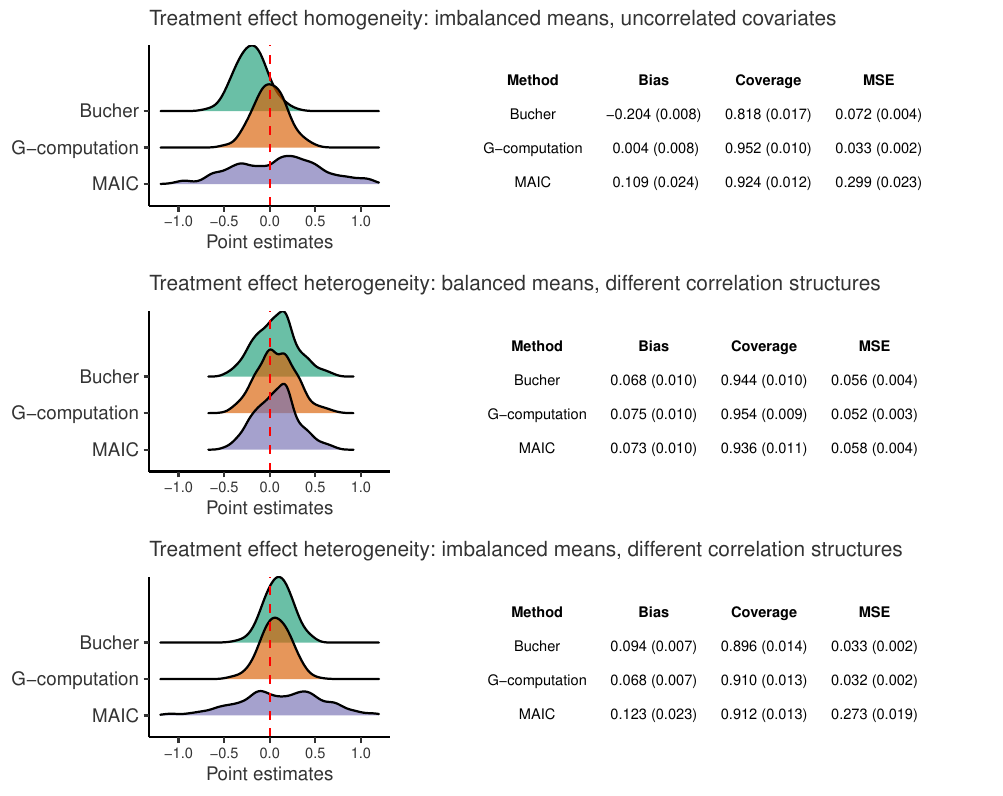}}
\caption[]
{
\textbf{Logistic outcome model and log odds ratio as the summary effect measure}. Point estimates of the marginal log odds ratio for $A$ versus $B$, and performance metrics with MCSEs for all methods across different settings.
}
\label{fig3}
\end{figure}

For the logistic outcome model with the log odds ratio as the summary measure (Figure \ref{fig3}), the performance of the Bucher method is deficient in the absence of treatment effect heterogeneity at the individual level. In this setting, the method displays substantial bias (-0.204), producing the highest absolute bias of all methods. Notice that the bias is virtually equal to the difference between the true marginal log odds ratios for $A$ versus $C$ in $S=1$ and $S=2$ ($\ln{2} - \ln{2.45}=-0.203$). The Bucher method also displays substantial undercoverage (empirical coverage rate of 0.818), resulting from the magnitude of the bias and the over-precision of standard errors. 

In the presence of treatment effect homogeneity at the individual level, the covariate adjustment methods improve performance with respect to the unadjusted approach. Parametric G-computation yields minimal bias (0.004) and an excellent empirical coverage rate (0.954), both within Monte Carlo error of the true marginal log odds ratio and the desired nominal 0.95 value, respectively. In this setting, MAIC exhibits bias (0.109) and displays slight undercoverage (empirical coverage rate of 0.924). Bias likely arises because the marginal log odds ratio depends on the full joint covariate distribution, even in the absence of treatment effect heterogeneity at the individual level. Enforcing cross-study balance between the marginal covariate moments does not necessarily guarantee balance in the joint covariate distributions after weighting. 

\clearpage

In the presence of individual-level treatment effect heterogeneity and differences between correlation structures, all methods produce comparable bias in the setting with balanced covariate means. The bias is 0.068 for the Bucher method, 0.075 for parametric G-computation and 0.073 for MAIC. In this setting, the level of bias is insufficient to degrade the empirical coverage rates (0.944 for the Bucher method, 0.954 for parametric G-computation and 0.936 for MAIC), which are not statistically significantly different to the desired 0.95 value, given 500 independent simulations. 

All methods present bias in the ``treatment effect heterogeneity'' setting with imbalanced marginal moments. Notably, the covariate adjustment approaches exhibit some bias despite accounting for all imbalances in covariate means. Performance improvements, if any, with respect to the unadjusted approach are limited. The bias is 0.068 for parametric G-computation and 0.123 for MAIC, compared to 0.094 for the Bucher method. While MAIC is the most biased method in this setting, it shows the least undercoverage, with an empirical coverage rate of 0.912 (compared to 0.896 for the Bucher method and 0.910 for parametric G-computation). In the case of the Bucher method and parametric G-computation, undercoverage seems driven by overly precise variance estimation. Conversely, undercoverage for MAIC appears to be bias-induced.

Parametric G-computation achieves greater precision and efficiency than MAIC, particularly in the settings with imbalanced covariate means. In these settings, MAIC is sensitive to poor overlap between the covariate distributions in $S=1$ and $S=2$. This inflates the variability of point estimates and the MSE. Again, the comparative imprecision of MAIC should not necessarily be viewed as an undesirable feature. Arguably, MAIC provides more “honest” uncertainty quantification by accounting for covariate differences while avoiding model-based extrapolation. 

Generally, we can conclude that, even in the absence of treatment effect heterogeneity at the individual level, unadjusted indirect comparisons are subject to bias. Covariate adjustment seems warranted where there are imbalances in marginal covariate moments, even if the covariates are purely prognostic. Where there are cross-study differences in correlation structures, covariate adjustment approaches that only account for differences in marginal moments are inherently limited in their ability to reduce bias. Once again, accounting for imbalances in correlations --- more generally, in the full joint covariate distributions -- seems necessary to remove bias.

\section{Discussion: implications for guidance}\label{sec5}

The results of the simulation study have key implications for current evidence synthesis guidance. When discussing ``current guidance'', we focus on recommendations provided by: (1) technical support documents from the National Institute for Health and Care Excellence (NICE) Decision Support Unit on heterogeneity, bias adjustment and meta-regression,\cite{dias2013evidencedos, dias2011nice} and on covariate-adjusted indirect comparisons with limited patient-level data;\cite{phillippo2018methods, phillippo2016nice} (2) reports by task forces of the International Society for Pharmacoeconomics and Outcomes Research on good research practices for indirect comparisons and network meta-analyses;\cite{jansen2011interpreting,jansen2014indirect,hoaglin2011conducting} (3) reviews and simulation studies on covariate-adjusted indirect comparisons authored by myself\cite{remiro2022parametric, remiro2021methods, remiro2022effect} and by others;\cite{phillippo2020assessing, petto2019alternative, weber2020comparison, kuhnast2017evaluation, jiang2020performance, cheng2020statistical} and (4) further guidance for evidence synthesis provided by influential research articles.\cite{jansen2012meta, cooper2009addressing, jansen2012directed, jansen2012network, jansen2008bayesian, coory2010frequency}

The literature makes a clear distinction between effect measure modifiers and prognostic variables.\cite{jansen2014indirect, phillippo2018methods, phillippo2016nice, remiro2022parametric, remiro2021methods, petto2019alternative,jiang2020performance} Effect measure modifiers are described as covariates inducing treatment effect heterogeneity at the individual level by ``interacting'' with treatment in an outcome model parametrized at such level.\cite{jansen2011interpreting, phillippo2018methods, phillippo2016nice, remiro2022parametric, jansen2012meta, dias2013evidencedos, dias2011nice, remiro2021methods, phillippo2020assessing,  petto2019alternative, weber2020comparison, kuhnast2017evaluation, jiang2020performance, cheng2020statistical, jansen2012network} According to the terminology used in Section \ref{sec2}, these would be conditional effect measure modifiers. As per Section \ref{sec2}, prognostic variables are conceptualized as covariates with main effects in the hypothetical outcome-generating model.\cite{phillippo2018methods, phillippo2016nice, remiro2022parametric, remiro2021methods, phillippo2020assessing, petto2019alternative, cheng2020statistical} Outcome predictors that do not induce a change in treatment response at the individual level are considered to be purely prognostic.\cite{phillippo2018methods, phillippo2016nice,remiro2022parametric, remiro2021methods}

\subsection{Unadjusted anchored indirect comparisons}\label{subsec51}

Unadjusted anchored indirect comparisons are known to be biased when there are cross-study imbalances in effect measure modifiers that interact with treatment.\cite{bucher1997results, jansen2011interpreting, phillippo2018methods, phillippo2016nice, jansen2012directed, cooper2009addressing, hoaglin2011conducting, remiro2021methods, jansen2008bayesian, coory2010frequency} Current guidance deems covariate adjustment to be unnecessary in the absence of treatment-covariate interactions, or in the rare instance in which the distributions of (conditional) effect measure modifiers are balanced between studies.\cite{jansen2011interpreting, phillippo2018methods, phillippo2016nice, jansen2012directed, remiro2021methods, remiro2022effect, kuhnast2017evaluation,coory2010frequency} For instance, Jansen et al. discourage adjusting for covariates when these are not (conditional) effect measure modifiers, stating that this may amplify bias in a meta-analysis.\cite{jansen2012directed} 

In the context of the simulation study in this article, current guidance would deem use of the Bucher method sensible in the setting with individual-level treatment effect homogeneity, because there are no treatment-covariate interactions in the outcome-generating models. According to many authors, randomization protects comparisons of relative effects from cross-study imbalances in purely prognostic covariates; these imbalances are not perceived to invalidate unadjusted anchored indirect comparisons.\cite{jansen2011interpreting,jansen2014indirect, phillippo2018methods, phillippo2016nice, dias2013evidencedos, dias2011nice, jansen2012directed, remiro2021methods, kuhnast2017evaluation} Nevertheless, where the target of inference is marginal and there is treatment effect homogeneity at the individual level, this is only the case for collapsible measures such as the mean difference and the (log) risk ratio.  

When comparing marginal treatment effects across studies, the unadjusted approach relies on no cross-trial differences in the variables modifying the marginal measure. For non-collapsible measures such as the (log) odds ratio, purely prognostic covariates can act as marginal effect measure modifiers at the study level. Therefore, cross-study imbalances in these covariates can bias the unadjusted anchored indirect comparison, as evidenced by the results of the simulation study. This motivates the use of covariate adjustment to account for imbalances in purely prognostic covariates, even in the absence of treatment effect heterogeneity at the individual level. 

In the context of the simulation study in this article, current guidance would also consider use of the Bucher method adequate in the setting with treatment effect heterogeneity at the individual level and balanced covariate marginal moments. The marginal covariate distributions are identical across studies. As such, there is cross-study balance in the means and standard deviations of the (conditional) effect measure modifiers (and of the purely prognostic covariates). As evidenced by the results of the simulation study, such guidance only applies for the linear outcome-generating model with a directly collapsible summary measure. 

In the presence of treatment effect heterogeneity at the individual level, marginal effects for summary measures that are not directly collapsible, such as the (log) risk ratio and (log) odds ratio, generally depend on the full joint distribution of (conditional) effect measure modifiers and purely prognostic covariates. Therefore, even if marginal covariate moments such as means and standard deviations -- for all (conditional) effect measure modifiers and prognostic variables -- are perfectly balanced across studies, unadjusted anchored indirect comparisons can still be subject to bias.  

Another limitation of the Bucher method is that, generally, it is over-precise in the simulation settings with non-linear models and summary measures that are not directly collapsible. This may be the case, even in scenarios where the method is unbiased, e.g.~log-linear outcome model with the log risk ratio as the summary measure and individual-level treatment effect homogeneity. Moreover, the extent of over-precision apparently depends on the sample size of the index study. For instance, in a simulation study by Remiro-Az\'ocar et al. (logistic outcome model, log odds ratio as the summary measure, individual-level treatment effect heterogeneity and imbalanced covariate means), variance underestimation tends to become more problematic as the number of subjects in the index trial increases.\cite{remiro2021methods}  

Given the shortcomings of the Bucher method, it is remarkable that certain HTA agencies, such as the Institute for Quality and Efficiency in Healthcare in Germany, only accept unadjusted indirect comparisons in the anchored scenario with a common comparator.\cite{iqwig2022general} Undoubtedly, these rely on a very stringent assumption: the unconditional constancy or transportability of marginal treatment effects between studies. The development of covariate adjustment methods that relax such assumption is imperative, even if these come with increased ``researcher degrees of freedom''.  

\subsection{Covariate-adjusted anchored indirect comparisons: variable selection}\label{subsec52}

In covariate-adjusted anchored indirect comparisons, the constancy assumption is conditional on a set of baseline covariates. Given adjustment for these covariates, the relative treatment effect is assumed constant across studies. While this assumption is strong, it is weaker than the unconditional constancy assumption required by the Bucher method. When transporting relative treatment effects, the consensus in the literature is that only ``effect modifiers'' need to be accounted for.\cite{jansen2014indirect, jansen2012directed, degtiar2023review, cole2010generalizing, zhang2016new, o2014generalizing, vo2023cautionary} 

\subsubsection{Weighting-based methods}\label{subsec521}

According to recent recommendations on weighting-based methods such as MAIC, for anchored indirect comparisons, only (conditional) effect measure modifiers for $A$ versus $C$ should be balanced to achieve unbiasedness in $S=2$.\cite{phillippo2018methods, phillippo2016nice, remiro2021methods, remiro2022effect, weber2020comparison} Available guidance asserts that no purely prognostic variables should be adjusted for, to alleviate the loss of effective sample size and precision after weighting.\cite{phillippo2018methods, phillippo2016nice, remiro2021methods, weber2020comparison} The guidance only appears warranted for collapsible measures in the absence of individual-level treatment effect heterogeneity, and for directly collapsible measures in the presence of such heterogeneity. 

Ultimately, weighting-based approaches such as MAIC target indirect comparisons of marginal treatment effects.\cite{remiro2021conflating} Consequently, all marginal effect measure modifiers should be balanced for unbiased estimation. Measures that are collapsible but not directly collapsible, such as the (log) risk ratio, generally depend on the full joint distribution of covariates, including those that are purely prognostic, where there is individual-level treatment effect heterogeneity. As such, purely prognostic covariates may modify the marginal treatment effect. With non-collapsible measures such as the (log) odds ratio, purely prognostic covariates can modify the marginal effect, even in the absence of treatment effect heterogeneity at the individual level.

Therefore, for the (log) risk ratio and the (log) odds ratio, cross-study imbalances in purely prognostic covariates may violate the conditional constancy assumption for the marginal effect. Bias can still be present if purely prognostic covariates are omitted by the analyst or unavailable in any of the studies. 

Two recent simulation studies co-authored by myself involve non-collapsible measures and anchored indirect comparisons of marginal effects.\cite{remiro2022parametric,remiro2021methods} In the corresponding simulations, four covariates are generated, all of which have imbalanced means but equal variances, marginal distribution forms and correlation structures. Two of the covariates are (conditional) effect measure modifiers, which induce treatment effect heterogeneity at the individual level through treatment-covariate interaction terms. The other two are purely prognostic variables contributing only main effects to the outcome-generating models. 

Interestingly, MAIC remains unbiased in both of the simulation studies despite not accounting for mean imbalances in the two purely prognostic variables. This is likely due to the interaction coefficients being relatively strong across the simulation scenarios.\footnote{Alternatively, Campbell et al. have hypothesized that this is due to the purely prognostic variables being imbalanced in terms of means but balanced in terms of variances.\cite{campbell2023standardization} According to their simulation study, cross-study differences in the variances of purely prognostic covariates appear to be more consequential than differences in their means for the transportability of the marginal (log) odds ratio.} Seemingly, in the presence of considerable effect measure modification at the individual level, the extent of marginal effect measure modification appears to be driven entirely by imbalances in the conditional effect measure modifiers. As illustrated by the present article, this is not necessarily the case in general. 

Having adopted model-based definitions of estimands and effect measure modification, the subset of covariates required to satisfy the conditional constancy assumption is at least as large for the marginal than for the conditional effect measure. Where the summary measure is not directly collapsible, the prospect of balancing additional covariates, beyond the (conditional) effect measure modifiers, is uncomfortable. Firstly, the level of covariate overlap will decrease as a larger number of covariates is selected. Under poor covariate overlap, the precision of weighting-based methods suffers due to extreme reductions in effective sample size.\cite{phillippo2020assessing} Therefore, larger trials would be necessary to mitigate precision losses and achieve efficient bias adjustment. 

Secondly, we cannot rule out that any of the purely prognostic covariates modifies the marginal effect measure. It is well known that variables that are prognostic of outcome are, almost invariably, (conditional) effect measure modifiers on at least one scale.\cite{lesko2018considerations} We now learn that the effect modifier status of a variable is not only defined with respect to a specific summary measure, but also dependent on whether such measure is marginal or conditional. For measures that are not directly collapsible, the assessment of effect modifier status on the marginal scale seems particularly challenging. Any variable that is associated with the outcome is potentially a marginal effect measure modifier. One can envision a situation where the analyst balances as many outcome predictors as possible to attempt meeting the conditional constancy assumption, thereby further inflating the variance of weighting-based methods. 

\subsubsection{Outcome modeling-based methods}\label{subsec522}

We now focus on covariate adjustment using regression models for the conditional outcome expectation. This includes methods for anchored pairwise indirect comparisons\cite{phillippo2018methods, phillippo2016nice, remiro2022parametric, remiro2021methods} and also meta-regression approaches,\cite{phillippo2020multilevel, jansen2012meta, cooper2009addressing, dias2013evidencedos, dias2011nice} which can handle larger networks of treatments and studies. In evidence synthesis, outcome models have typically targeted a conditional estimand, e.g.~the treatment effect estimate is given by the treatment coefficient of a multivariable regression parametrized at the individual level.\cite{phillippo2018methods, phillippo2016nice, phillippo2020multilevel, jansen2012meta, cooper2009addressing, dias2013evidencedos, dias2011nice, remiro2021conflating, remiro2021methods,phillippo2020assessing, jansen2012network} 

As a result, recommendations are oriented towards estimating well a conditional effect and have placed focus on modeling treatment-covariate interactions.\cite{jansen2011interpreting, phillippo2018methods, phillippo2016nice, phillippo2020multilevel, jansen2012meta, cooper2009addressing, dias2013evidencedos, dias2011nice,  hoaglin2011conducting, remiro2021methods, phillippo2020assessing, weber2020comparison, jansen2012network} In the outcome model, only the inclusion of imbalanced (conditional) effect measure modifiers is deemed necessary to reduce bias.\cite{jansen2011interpreting, jansen2014indirect, phillippo2018methods, phillippo2016nice, remiro2021methods} The inclusion of balanced (conditional) effect measure modifiers and purely prognostic variables is considered optional.\cite{jansen2014indirect, phillippo2018methods, phillippo2016nice, remiro2021methods} It is not believed to remove bias further but is encouraged if the fit of the outcome model improves, leading to more precise estimation of the conditional treatment effect.\cite{phillippo2018methods, phillippo2016nice, remiro2021methods} 

To target marginal effects, outcome modeling-based methods must be extended using model-based G-computation or standardization approaches.\cite{remiro2022parametric, phillippo2021target} These predict counterfactual outcomes under each treatment by applying the fitted regression to the sample or population of interest. The marginal effect is derived from marginal mean outcome predictions for each treatment, which, in turn, are derived from subject-specific outcome predictions.\cite{remiro2022parametric} 

Therefore, reliable predictions of absolute outcomes at the individual level will be required. Generally, correct specification of the outcome model is necessary for unbiased estimation of the marginal effect. Namely, the conditional constancy assumption for the marginal effect is enforced by satisfying the conditional constancy assumption for absolute outcomes.\cite{josey2021transporting} Certainly with non-linear models, the outcome model should account for (conditional) effect measure modifiers that are balanced prior to adjustment, and for purely prognostic variables as well. While adjusting for additional outcome predictors increases the variance of weighting-based methods, it should decrease the variance of outcome modeling-based estimators. 

An interesting corollary is that there is more overlap than previously thought between the assumptions made by covariate-adjusted anchored indirect comparisons and those made by their unanchored counterparts without a common comparator. For instance, according to a NICE Decision Support Unit technical support document, unanchored comparisons assume that ``absolute outcomes can be predicted from the covariates'' and that ``all (conditional) effect (measure) modifiers and prognostic factors are accounted for and correctly specified''.\cite{phillippo2016nice} The document states that these assumptions are largely considered to be ``implausibly strong'' and ``impossible to meet''.\cite{phillippo2016nice} Nevertheless, such assumptions may also be required when comparing marginal effects in the anchored scenario.  

\subsection{Covariate-adjusted anchored indirect comparisons: distributional assumptions}\label{subsec53}

In practice, individual patient data for $S=2$ are often unavailable. Only summary moments for the marginal covariate distributions are reported in publications, and information on the full joint covariate distribution, e.g.~distributional forms and correlation structure, is rarely available. Consequently, covariate-adjusted indirect comparisons rely on unverifiable assumptions to approximate the joint covariate distribution in the target. As stated by Phillippo et al., ``further research is needed to investigate the extent of error following from the availability of only marginal, rather than joint, covariate distributions''.\cite{phillippo2016nice} 

Weighting-based approaches such as MAIC and outcome modeling-based methods such as parametric G-computation make slightly different covariate-distributional assumptions. Those of the former are more implicit and nuanced, whereas those of the latter are more explicit. In any case, the assumptions matter. Marginal estimands for certain effect measures do not only depend on marginal covariate moments, but on the full joint distribution of covariates, even if these are purely prognostic and do not directly induce treatment effect heterogeneity at the individual level. As shown in the ``heterogeneity'' settings of the simulation study in this article, ignoring cross-study differences in covariate correlations may compromise the constancy of marginal effects and lead to bias. 

\subsubsection{Weighting-based methods}\label{subsec531}

Due to the lack of published correlation information for $S=2$, weighting methods based on aggregate-level data such as MAIC can only attain cross-study balance in marginal covariate moments. However, such balance does not guarantee multidimensional balance across the joint covariate distributions. As stated in the NICE Decision Support Unit technical support document on covariate-adjusted indirect comparisons, ``when covariate correlations are not available from the ($S=2$) population, and therefore cannot be balanced by inclusion in the weighting model, they are assumed to be equal to the correlations amongst covariates in the pseudo-population formed by weighting the ($S=1$) population''.\cite{phillippo2016nice}

The simulation study in this article examines the adequacy of only balancing marginal moments such as means, while ignoring differences in correlations. Whether marginal covariate balance suffices or not depends on the outcome-generating model and summary effect measure. Under the linear outcome-generating models in the simulation study, with the marginal mean difference as the targeted effect measure, mean-balancing weights perform adequately in all settings. In the ``treatment effect homogeneity'' setting, weighting is not even necessary because the marginal mean difference is constant across studies. In the ``treatment effect heterogeneity'' settings, mean-balancing suffices to remove bias because the conditional mean difference is linear on the (conditional) effect measure modifier.\cite{josey2021transporting, cheng2023double} As the mean difference is directly collapsible, the marginal treatment effect is linear on the mean of the (conditional) effect measure modifier, as illustrated in Section \ref{subsec23}. 

Conversely, if the conditional mean difference were to depend on non-trivial transformations, e.g.~higher-order powers, covariate-by-covariate interactions, flexible basis functions or non-linear functions, of the covariates, mean-balancing weights would not guarantee bias removal. For instance, if the outcome-generating model contains a squared covariate-by-treatment product term, mean-balancing can incur bias if only first-order moments (means) and not second-order moments (variances) are balanced. Because second-order balance is enforced by balancing the means of squared covariates,\cite{phillippo2016nice} balancing both first- and second-order moments would provide some protection against bias.\cite{campbell2023standardization} Interestingly, certain authors have recommended against this, likely because it reduces the effective sample size after weighting and the precision of the treatment effect estimate.\cite{weber2020comparison, hatswell2020effects}

Under the log-linear outcome-generating models in the simulation study, with the marginal log risk ratio as the targeted effect measure, the situation is more complex. In the ``treatment effect homogeneity'' setting, weighting is not necessary because the (log) risk ratio is collapsible and the marginal measure is constant across studies. However, mean-balancing weights perform sub-optimally in the ``treatment effect heterogeneity'' settings. Even though the conditional log risk ratio is linear on the (conditional) effect measure modifier, mean-balancing is insufficient to remove bias because correlation structures differ between studies and the marginal log risk ratio depends on the full joint covariate distribution.

Under the logistic outcome-generating models in the simulation study, with the marginal log odds ratio as the targeted effect measure, mean-balancing weights do not perform adequately in any setting. Even in the absence of individual-level treatment effect heterogeneity, enforcing mean balance is insufficient to remove bias, as it does not necessarily guarantee balance in the joint covariate distributions after weighting. Because the (log) odds ratio is non-collapsible, the marginal measure depends on the full joint covariate distribution, including that of purely prognostic covariates, in all settings. 

For all effect measures, but particularly for those that are not directly collapsible, imposing balancing constraints on additional summary statistics is expected to increase bias-robustness. However, balancing correlations, central moments of higher order than variances or complex covariate transformations is not possible based on typical reporting requirements for the $S=2$ publication. 

Moreover, for entropy balancing techniques such as MAIC, there is always a tension between satisfying the conditional constancy of marginal effects and there being a solution to the convex optimization problem. Imposing a greater number of balancing constraints increases the plausibility of the former but decreases the likelihood of the latter. If the number of constraints is too high, MAIC may suffer from convergence failures and not even produce an estimate.\cite{phillippo2020assessing} If a feasible solution to the optimization problem does exist, the trade-off is between satisfying the conditional constancy assumption and being able to maintain a reasonable level of precision.\cite{campbell2023standardization} Increasing the number of balancing constraints will have a cost: further effective sample size reductions after weighting and wider interval estimates.\cite{campbell2023standardization}  

\subsubsection{Outcome modeling-based methods}\label{subsec532}

With no individual patient data for $S=2$, outcome modeling-based methods assume that the joint covariate distribution of the target has been characterized correctly, by the combination of specified marginal distribution forms and correlation structure. In the simulation study in this article, G-computation assumes that the parametric forms of the marginal distributions and pairwise linear correlations in $S=2$ are equal to those observed for $S=1$. 

Following recommendations in the literature,\cite{remiro2022parametric, phillippo2020multilevel, phillippo2020assessing, ishak2015simulation} we have decided to mimic the pairwise correlations of the $S=1$ covariates, which are uncorrelated in expectation. The rationale behind such recommendations is that the relationships between covariates should remain similar across trials. Nevertheless, this is arguably an unrealistic assumption. Covariate correlation structures are likely to differ between studies with different selection criteria, sampling or recruitment mechanisms. 

The ``treatment effect heterogeneity'' settings of the simulation study in this article investigate whether the performance of parametric G-computation is sensitive to the ``equal correlations'' assumption where there are cross-study differences in correlation structures. For the linear outcome model with the mean difference as the summary effect measure, parametric G-computation remains unbiased. This is because the mean difference is directly collapsible and the outcome-generating model is a relatively simple model, with only one first-order (two-way) treatment-covariate interaction. Consequently, the marginal mean difference only varies with the (conditional) effect measure modifier mean and does not vary with the covariate correlations. 


Conversely, the ``equal correlations'' assumption is susceptible to bias where the marginal effect measure depends on the full joint covariate distribution and there are cross-study differences in correlation structures. That is, for the log-linear outcome model with the log risk ratio as the summary effect measure, and for the logistic outcome model with the log odds ratio. Assuming that there are cross-study differences in correlations, the ``equal correlations'' assumption is expected to be problematic for the log odds ratio, but not the log risk ratio, in the absence of individual-level treatment effect heterogeneity and where all covariates are purely prognostic. 

A recent simulation study concludes that the aforementioned covariate-distributional assumptions have negligible impact on the performance of outcome modeling-based methods, both in terms of bias and variance estimation, even if the assumptions are incorrect.\cite{phillippo2020assessing} The cited simulation study features a logistic outcome model and the log odds ratio as the summary measure. Its target estimand is a model-based conditional treatment effect, which may explain why its conclusions conflict with ours. 


There are other recommendations in the literature that do not apply to anchored indirect comparisons of marginal effects. For instance, that the misspecification of correlations in the target will only incur bias if the outcome-generating model contains treatment-covariate interactions of second-order or higher.\cite{phillippo2016nice, remiro2021methods} Also, that the misspecification of correlations involving purely prognostic variables does not incur bias due to the cancellation of terms.\cite{phillippo2016nice, remiro2021methods} That such recommendations are not applicable where the target estimand is marginal is evidenced by the simulation study in this article. Here, the marginal log risk ratio and log odds ratio can differ across studies due to correlations involving purely prognostic covariates, and in the absence of treatment-covariate interactions that are second-order or higher. 

The simulation study in this article does not explore the implications of incorrectly specifying the marginal covariate variances and distributional forms in the target, because these are identical across studies in the covariate-generating mechanisms. The impact of failures in these assumptions should be investigated further in future simulation studies.  

\subsection{Assessment of statistical interactions}\label{subsec54}

Effect measure modifiers are often identified by evaluating treatment-covariate interaction terms in regression models fitted to individual patient data.\cite{greenland1983tests, brookes2004subgroup} So-called interaction tests are demanded by many HTA agencies, such as the Institute for Quality and Efficiency in Healthcare in Germany, to assess external validity.\cite{iqwig2022general} A well-established issue is that, in general, RCTs are severely underpowered to detect interactions via statistical testing.\cite{greenland1983tests, brookes2004subgroup} 

In addition to this, marginal effect measure modification may occur in the absence of individual-level treatment effect heterogeneity where there is non-collapsibility. Therefore, even if trials were sized large enough to detect treatment-covariate interactions, variables should not be discarded for adjustment on the grounds of large $p$-values. It may still be reasonable to adjust for a covariate when the null hypothesis of homogeneity is not rejected. With measures that are not directly collapsible, this may be necessary to avoid bias in indirect comparisons of marginal treatment effects.

Current guidance on anchored indirect comparisons suggests presenting quantitative evidence on the potential bias removal that is to be incurred with covariate adjustment compared to the unadjusted approach. Phillippo et al. state that a covariate-adjusted analysis ``should only be submitted if, putting together the magnitude of the supposed interaction with the extent of the imbalance (in covariates between the studies), a material difference in the estimated treatment comparisons would be obtained''.\cite{phillippo2016nice} Similarly, Remiro-Az\'ocar et al. assert that ``the interaction coefficient can be multiplied by the difference in (conditional) effect (measure) modifier means to gauge the level of induced bias''.\cite{remiro2021methods} 

Again, where the inferential target is a marginal effect and the summary measure is not directly collapsible, bias can be induced even if the (conditional) effect measure modifiers have balanced marginal moments. Where the summary measure is non-collapsible, bias can still be induced in the absence of treatment-covariate interactions. Covariate adjustment may still be warranted in these cases.

At times, none of the studies included in an evidence synthesis match the relevant target population for decision-making. In these cases, the shared (conditional) effect modifier assumption\cite{jansen2012meta, cooper2009addressing, dias2013evidencedos, dias2011nice, jansen2012network} is invoked to transport relative effect estimates for the active-active treatment comparison to any given target population.\cite{phillippo2018methods, phillippo2016nice, phillippo2020multilevel, phillippo2020assessing} This assumption implies that active treatments have the same set of (conditional) effect measure modifiers with respect to the common comparator, and that treatment-covariate interactions are identical for both treatments. This allows (conditional) effect measure modifiers to cancel out and conditional treatment effects for the active-active comparison to be applicable to any target population (given covariate adjustment, the conditional effect is constant across populations). 

As currently conceptualized, and contrary to prior assertions,\cite{remiro2021methods} the shared (conditional) effect modifier assumption does not necessarily allow for transporting marginal effects across populations. Consider the ``treatment effect homogeneity'' setting with the logistic outcome model in the simulation study in this article. Even though the true conditional log odds ratio for the active-active treatment comparison is constant (zero) across all subjects in both studies, the true marginal log odds ratio differs between $S=1$ and $S=2$. 

\subsection{Transportability}\label{subsec55}

When making population-level decisions in HTA, the scientific question translates into a marginal estimand.\cite{remiro2022target} We have assumed that the inferential target is a marginal effect. Inevitably, health technology assessors and stakeholders will attempt to generalize effect estimates to the relevant population for policy-making. Covariate-adjusted indirect comparisons and network meta-regressions attempt to do this explicitly. Other evidence synthesis methods do not, but still invoke a constancy assumption. 

Consequently, transportability is a central property to consider for the choice of a suitable effect measure.\cite{liu2022correct, martinussen2013collapsibility, xiao2022controversy, didelez2022logic, huitfeldt2021shall} For different types of summary measures, different classes of covariates will compromise the transportability of marginal treatment effects.\cite{webster2023choice} For measures that are not directly collapsible, the dependence of marginal effects on the distribution of prognostic factors, even if these are not determinants of treatment response at the individual level, suggests that such summary measures are not appealing for transportability.\cite{xiao2022controversy, didelez2022logic, huitfeldt2021shall} 

Consider that the outcome is binary. Common summary measures for the treatment effect would be the risk difference (directly collapsible), the (log) risk ratio (collapsible but not directly collapsible) and the (log) odds ratio (non-collapsible).\cite{colnet2023risk, webster2023choice} Let's assume that there is treatment effect heterogeneity at the individual level, as assuming otherwise is arguably an over-simplification of a complex reality. In this case, the marginal risk difference depends on the distribution of (conditional) effect measure modifiers, the marginal (log) risk ratio depends on the joint distribution of (conditional) effect measure modifiers and purely prognostic covariates that are associated with the former, and the marginal (log) odds ratio depends on the joint distribution of (conditional) effect measure modifiers and purely prognostic covariates, even if these are not associated with the former. 

As stated by Webster-Clark and Keil, this does not necessarily imply that the marginal risk difference requires a smaller set of covariates to account for, because covariates may not modify conditional treatment effects on all measurement scales.\cite{webster2023choice} Nevertheless, if all candidate covariates are prognostic of outcome, transporting the marginal (log) odds ratio will always require accounting for the greatest number of covariates.\cite{webster2023choice} 

As highlighted in Section \ref{subsec521} and Section \ref{subsec54}, whether an effect measure is collapsible or not, and whether it is directly collapsible or not, has implications on screening for effect measure modification on the marginal scale. The lack of direct collapsibility complicates the selection of baseline characteristics for covariate adjustment. This process is typically guided by biological rationale, clinical expert judgement and subject matter knowledge about determinants of treatment response at the individual level.\cite{huitfeldt2021shall} With measures that are not directly collapsible, variables that are not determinants of individual-level treatment response can modify marginal treatment effects. For non-collapsible measures, this is the case even in the total absence of treatment effect heterogeneity at the individual level.  

Meta-analyses routinely pool non-collapsible effect measures such as log odds ratios for binary outcomes and log hazard ratios for time-to-event outcomes.\cite{dias2013evidence, lumley2002network, lu2004combination, phillippo2018methods, phillippo2016nice, remiro2022parametric, cooper2009addressing, dias2013evidencedos, dias2011nice, jansen2012directed, hoaglin2011conducting, remiro2021methods, remiro2022effect, jansen2012network} The popularity of (log) odds and (log) hazard ratios largely stems from their symmetry and from the attractive statistical properties of logistic and Cox proportional hazards regression, respectively. For binary outcomes, logistic regression guarantees that fitted outcome probabilities lie in the $(0, 1)$ interval.\cite{localio2007relative} It is worth noting that use of a logistic regression for covariate adjustment does not necessarily imply that the target estimand is a (log) odds ratio. Through model-based standardization, one can compute marginal treatment effects on collapsible scales from the probabilities predicted by the fitted regression.\cite{localio2007relative, greenland2004model} The selected collapsible scale for the target estimand would be used to define and test for (conditional) effect measure modification or interaction, which are scale-specific. 

From our exposition, readers may conclude that conditional treatment effects are more transportable than marginal treatment effects. This is a position sometimes taken by authors in what is an active area of debate.\cite{phillippo2021target, remiro2022target, spieker2022comments, van2022estimands} It may be a valid position: the conditional estimands in the simulation study can be expressed in relatively simple terms, while the marginal estimands are complex expressions that can depend on the full joint covariate distribution where the summary measure is not directly collapsible. There is a caveat: such properties are an artefact of adopting ``model-based'' estimand definitions and correct statistical assumptions about model specification. They do not intrinsically apply to ``model-free'' estimands. In practice, there is no guarantee that modeling assumptions will hold, and the ``true'' outcome-generating model will be more complicated than those proposed in this article. 

Finally, it is worth noting that the the terms ``marginal'' and ``conditional'' are inherently relative when synthesizing evidence across different studies. For instance, the $SATE$ in Equation \ref{additive_study} is a marginal treatment effect within $S=1$ and the $TATE$ in Equation \ref{additive_target} is a marginal treatment effect within $S=2$. We use the term ``marginal'' because the estimand of interest refers to how the \textit{within-study} marginal distribution of the outcome varies with a change in treatment. If one were to combine studies $S=1$ and $S=2$, each study could be viewed as a different subgroup, and the within-study marginal estimands would become conditional on study membership.\cite{van2022estimands, vo2021assessing} 

\section{Concluding remarks}\label{sec6}

Evidence synthesis in connected networks typically relies on the constancy of relative treatment effects between studies. When adjusting for covariates, this assumption is conditional on a set of baseline characteristics. Current guidance establishes that, for constancy to hold, there is either: (1) no effect measure modification by the covariates; or (2) the effect measure modifiers are equidistributed across studies. In evidence synthesis, effect measure modifiers have traditionally been described as covariates that induce treatment effect heterogeneity at the individual level, through treatment-covariate interactions in an outcome model parametrized at such level. Therefore, effect modification has been defined with respect to a conditional measure, even though the relevant target estimand for population-level decisions in HTA is a marginal effect. 

For certain summary measures, the set of marginal and conditional effect measure modifiers may not coincide. In the absence of individual-level treatment effect heterogeneity, marginal effects for non-collapsible measures such as the (log) odds ratio generally depend on the distribution of purely prognostic covariates that are not effect measure modifiers on the conditional scale. In the presence of individual-level treatment effect heterogeneity, marginal effects for measures that are not directly collapsible, such as the (log) risk ratio and the (log) odds ratio, generally depend on the full joint distribution of purely prognostic covariates and covariates that are effect measure modifiers on the conditional scale. 

On the marginal scale, depending on the mathematical properties of the selected summary measure, different types of covariates must be accounted for to achieve external validity or transportability with respect to a given target population. Namely, the types of covariates classed as marginal effect measure modifiers are a function of the summary measure. 

Collapsible effect measures are appealing for transportability because they remove dependence on model-based covariate adjustment where there is treatment effect homogeneity at the individual level. In this setting, marginal effects for collapsible measures do not depend on the distribution of purely prognostic covariates and do not vary across studies. Directly collapsible effect measures are appealing for transportability because they can reduce dependence on model-based covariate adjustment, either where there is treatment effect homogeneity at the individual level, or where there is heterogeneity and marginal covariate moments are balanced across studies. Moreover, direct collapsibility facilitates the selection of baseline covariates for adjustment where there is treatment effect heterogeneity at the individual level.  

Questions are raised about the performance of covariate-adjusted indirect comparisons in the absence of individual patient data for the target. Marginal estimands for measures that are not directly collapsible depend on the full joint covariate distribution, not only on marginal covariate moments. Where there are cross-study differences in correlation structures, methods that only account for differences in marginal moments are inherently limited in their ability to remove bias. Accounting for cross-study differences in correlations -- more generally, in the full joint covariate distributions -- appears necessary to improve performance. 

While this article addresses heterogeneity (variation in the same treatment contrast across studies), it does not address inconsistency (discrepancies between direct and indirect comparisons in a network of studies). The concepts explored in this article are also relevant to the latter, which arises from imbalances in effect measure modifiers across comparisons. In network meta-analyses of marginal effects, depending on the summary measure, inconsistency could emerge from imbalances in purely prognostic variables or correlation structures between the studies providing direct and indirect evidence. 

Finally, it is worth noting that the concept of “aggregation bias”, a form of ecological bias, has already been well-documented in evidence synthesis for covariate adjustment methods making use of aggregate-level data.\cite{jansen2012meta, saramago2012mixed, donegan2013combining, berlin2002individual} For instance, the literature highlights that meta-regression methods which assume common coefficients for individual-level and aggregate study-level covariates are susceptible to such type of bias. In this context, Jansen and Cope outline that “the association between a patient characteristic and the treatment effect of the studied interventions at the study level may not reflect the individual-level effect modification of that covariate''.\cite{jansen2012meta}

With the exception of a recent article by Riley et al.,\cite{riley2023using} the connection between aggregation bias and collapsibility was yet to be made explicit. Notably, much of the guidance cited in this article warns the reader about aggregation bias in the context of meta-regressions, but: (1) deems the standard unadjusted anchored indirect comparisons to be acceptable in the absence of individual-level treatment effect heterogeneity or treatment-covariate interactions, or where (conditional) effect measure modifiers are equidistributed across studies;\cite{jansen2011interpreting, phillippo2018methods, phillippo2016nice, cooper2009addressing, jansen2012directed, hoaglin2011conducting} (2) suggests that covariate adjustment is not warranted in such scenarios;\cite{jansen2011interpreting, phillippo2018methods, phillippo2016nice} or (3) indicates that it is not necessary to account for cross-study imbalances in purely prognostic variables in covariate-adjusted anchored indirect comparisons.\cite{jansen2011interpreting, phillippo2018methods, phillippo2016nice, dias2013evidencedos, dias2011nice} This article should help to establish some clarity. 

\section*{Acknowledgments}

This work was strongly motivated by the insightful feedback of Reviewer 1 to a previous article co-authored by myself,\cite{remiro2021methods} and considerably improved following valuable feedback from Tim Morris. I extend my sincere gratitude and appreciation to both. I thank the editor and anonymous peer-reviewers for their insightful comments, which have helped to improve the article further. 

\subsection*{Financial disclosure}

No funding to report. 

\subsection*{Conflict of interest}

The author is employed by Novo Nordisk. No conflicts of interest are declared as this research is purely methodological. 

\subsection*{Data Availability Statement}

The files required to generate the data, run the simulations, and reproduce the results are available at \url{http://github.com/remiroazocar/conditional_marginal_effect_modifiers}.

\normalsize
\bibliography{wileyNJD-AMA}

\end{document}